\documentclass[10pt,final,doublecolumn]{IEEEtran}
\usepackage{amsmath}
\usepackage{amssymb}
\usepackage{amsfonts}
\usepackage{latexsym}
\usepackage{graphicx}
\usepackage{bbding}
\usepackage{enumerate}
\usepackage{subeqnarray}
\usepackage{multicol}
\usepackage{color}
\usepackage{setspace}
\usepackage{mathrsfs}
\usepackage{array}
\usepackage{algorithm,algorithmic}
\usepackage{colortbl}
\usepackage{bm}
\IEEEoverridecommandlockouts
\allowdisplaybreaks[4]

\begin{document}
\title{Energy-Efficient Design of Satellite-Terrestrial Computing in 6G Wireless Networks}
\author{
Qi Wang, Xiaoming Chen, and Qiao Qi
\thanks{Qi Wang, Xiaoming Chen, and Qiao Qi are with the College of Information Science and Electronic Engineering, Zhejiang University, Hangzhou 310027, China (e-mails: \{wang-qi, chen\underline{~}xiaoming, and qiqiao1996\}@zju.edu.cn).}}\maketitle

\begin{abstract}
In this paper, we investigate the issue of satellite-terrestrial computing in the sixth generation (6G) wireless networks, where multiple terrestrial base stations (BSs) and low earth orbit (LEO) satellites collaboratively provide edge computing services to ground user equipments (GUEs) and space user equipments (SUEs) over the world. In particular, we design a complete process of satellite-terrestrial computing in terms of communication and computing according to the characteristics of 6G wireless networks. In order to minimize the weighted total energy consumption while ensuring delay requirements of computing tasks, an energy-efficient satellite-terrestrial computing algorithm is put forward by jointly optimizing offloading selection, beamforming design and resource allocation. Finally, both theoretical analysis and simulation results confirm fast convergence and superior performance of the proposed algorithm for satellite-terrestrial computing in 6G wireless networks.
\end{abstract}


\section{Introduction}
With the fast development of information technology in recent years, many new intelligent applications and services have emerged, such as extended reality, holographic communication, and autonomous driving which require mass data processing. However, due to limited computing power, it is impossible to complete mass data processing at the terminals in real time. In this context, mobile edge computing (MEC) has become a key enabling technology for the fifth generation (5G) wireless networks by deploying computing servers at the network edge, e.g., base station (BS), to provide low-latency computing services \cite{MEC}. However, 5G wireless networks only cover a small proportion of the world. According to statistics, more than half of the global region, especially in oceans, deserts and remote mountainous areas, still suffers from the lack of Internet access \cite{statistics}. Thus, it is desired to design the sixth generation (6G) wireless networks with ubiquitous communication and real-time computing capabilities.

On the one hand, for the special areas that the terrestrial communication cannot cover, satellite communication can be adopted as a supplement for providing access services. Compared to terrestrial communication, satellite communication has a large coverage area and a more flexible deployment due to the geographical advantages, which has been widely regarded as one of typical application scenarios of 6G wireless network \cite{6G satellite1}-\cite{6G satellite3}. In particular, low earth orbit (LEO) satellites with lower orbital altitude can provide high-efficiency communications because of the low transmission delay and propagation pathloss \cite{LEO1} \cite{LEO2}. As is well-known to all, SpaceX is executing the ``Starlink" program, which aims to launch 12,000 LEO satellites by 2024 for providing global satellite access \cite{spacex}. On the other hand, due to limited computing resources, MEC servers at the terrestrial BSs are often overloaded in computing-intensive areas. In this context, LEO satellites equipped with MEC servers can be regarded as space network nodes to offer additional computing power. Driven by these issues, it makes sense to explore satellite-terrestrial computing in 6G wireless networks to meet the requirements of high-reliable communication and low-latency computing for various intelligent applications around the world \cite{satellite-terrestrial network}.

Generally speaking, satellite-terrestrial computing makes use of global-covered integrated satellite-terrestrial network to provide edge computing \cite{related work 0}. For satellite-terrestrial computing, communication and computing are two key issues affecting the edge computing. The former decides the quality of data transmission, and the latter determines the performance of data processing. To improve the quality of data transmission for computing tasks, the communication in integrated satellite-terrestrial networks has been extensively studied. For instance, the authors investigated the channel characteristics of the integrated satellite-terrestrial system and designed a strategy to maximize the utilization of communication resources \cite{related work 1}. In \cite{add share spectral}, the transmission performance of a cognitive satellite-terrestrial system was analyzed when the satellite link and the terrestrial link shared the same spectrum. In order to avoid inter-user interference, orthogonal resources are usually allocated to the terminals in satellite-terrestrial systems, resulting in a low spectrum utility \cite{OMA1}, \cite{OMA2}. To cope with this issue, non-orthogonal multiple access (NOMA)-based satellite-terrestrial systems are proposed to improve the spectral efficiency. In \cite{related work 6} and \cite{related work 2}, the authors studied the beamforming design and power allocation to reduce the co-channel interference in NOMA-based satellite-terrestrial systems. Moreover, some novel communication techniques, such as rate splitting multiple access \cite{add RSMA}, reconfigurable intelligent surface-assisted \cite{add RIS}, and millimeter wave communication \cite{add millimeter wave}, have been extensively studied in research related to satellite-terrestrial networks.

For computing, it is desired to select an appropriate computing node, namely computing offloading, according to the characteristics of satellite-terrestrial computing. Previously, computing offloading in terrestrial networks has been well investigated \cite{TerrestrialComputing}. For example, a joint offloading decision and resource allocation scheme was provided in \cite{offloading 1} to minimize the total energy cost in the mobile cloud networks. Moreover, the authors in \cite{offloading 2} studied an energy-aware task offloading problem for user-intensive terrestrial systems. However, computing offloading for integrated satellite-terrestrial networks is still an open issue. This is due to the more complex transmission environment for the integrated satellite-terrestrial network, which requires a comprehensive architecture to coordinate wireless and computing resources across multiple terminals and nodes. To this end, a series of works have focused on resource allocation in satellite-terrestrial networks. For example, a novel double edge computing framework of satellite-terrestrial networks was proposed in \cite{related work 3} to minimize the offloading delay and the required energy by resource allocation. The authors in \cite{related work 4} jointly optimized task execution sequence and computing resource allocation for the satellite-terrestrial double edge computing system. In \cite{related work 5}, the authors put forward a joint offloading strategy and resource scheduling design to improve the overall performance for integrated satellite-terrestrial networks with hybrid cloud and edge computing.

Recently, there has been a growing emergence of satellites and space applications, including full-coverage communications, climate monitoring, navigation and positioning, earth observation and aeronautical exploration, which also require some certain computing power. Therefore, LEO satellites are expected to serve as space computing nodes to provide communication and computing services for space terminals. Nevertheless, existing works lack consideration of inter-satellite communication and space resource sharing, as well as the deep collaboration of multiple types of computing nodes in integrated satellite-terrestrial networks. In this context, this paper aims to design a general framework for satellite-terrestrial computing in 6G wireless networks, providing high-reliable communication and low-latency computing services for ground user equipments (GUEs) and space user equipments (SUEs) simultaneously. The contributions of this paper are as follows.

\begin{enumerate}
\item We present a general satellite-terrestrial computing framework for providing global seamless computing power supply. In particular, multiple types of computing nodes cooperatively provide low-latency computing power support to both GUEs and SUEs simultaneously. The framework meets the increasing demand for space applications and supplements terrestrial computing power.

\item Within satellite-terrestrial computing system, we evaluate the performance in terms of the energy consumption and execution time, which yields valuable insights for optimal design. To minimize the weighted total energy consumption while ensuring delay requirements, we formulate a key problem for satellite-terrestrial computing by jointly optimizing offloading selection, beamforming design and resource allocation.

\item  To solve the formulated complex optimization problem, we decompose the original NP-hard problem into three subproblems. For the offloading selection subproblem, we adopt the relaxation mapping method to strike a balance between optimization results and computational complexity. For the beamforming design and resource allocation subproblems, a combination of closed-form solutions and convex approximation techniques is employed for effective solution. The effectiveness and superiority of the proposed algorithm are validated through theoretical analysis and simulation results.
\end{enumerate}

The rest of this paper is organized as follows: Section II introduces the satellite-terrestrial computing model. Section III designs a satellite-terrestrial computing algorithm to minimize the weighted total energy consumption. Section IV presents simulation results to verify the effectiveness of the proposed algorithm. Finally, Section V concludes the paper.

\emph{Notations}: Bold lower case and upper case letters denote column vectors and matrices, respectively. $(\cdot)^H$, $\text{Rank}(\cdot)$ and $\text{tr}(\cdot)$ indicate the conjugate transpose, the rank and the trace of a matrix, respectively. $|\cdot|$ means the absolute value of a scalar, $\|\cdot\|$ means the 2-norm of a vector, $\mathbf{X}\succeq\mathbf{0}$ means that matrix $\mathbf{X}$ is a positive semi-definite matrix, ${{\mathbb{C}}^{a\times b}}$ denotes the set of $\emph{a}$ $\times$ $\emph{b}$ dimensional complex matrixes, and $\odot$ denotes Hadamard product.

\section{System Model}
\begin{figure}[h] \centering
\includegraphics [width=0.5\textwidth] {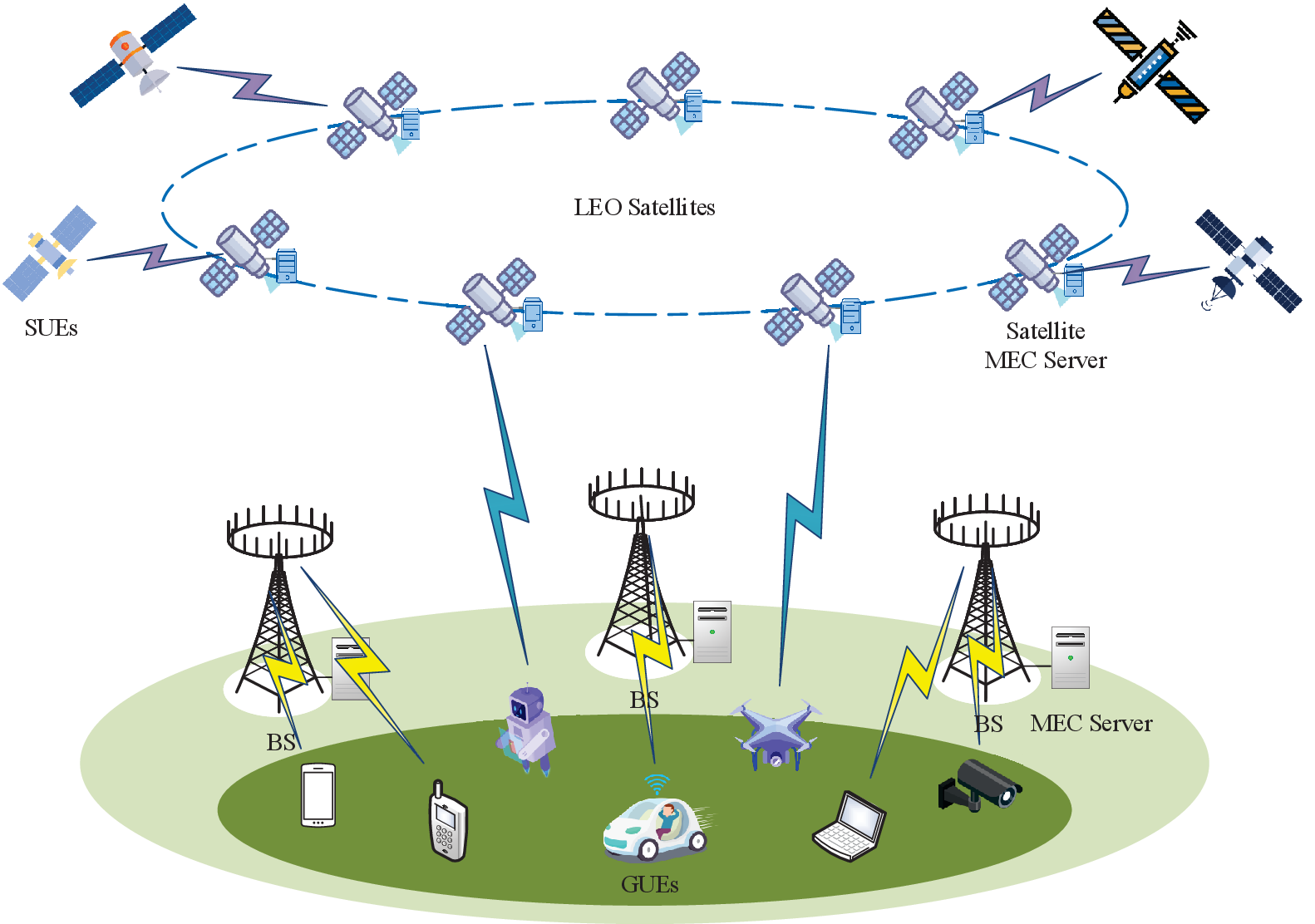}
\caption {System model for satellite-terrestrial computing in 6G wireless networks.}
\label{Fig1}
\end{figure}
Let us consider an integrated satellite-terrestrial 6G wireless network consisting of $M$ BSs each equipped with $N_t^g$ antennas, $N$ LEO satellites each equipped with $N_t^s$ antennas, $K$ single-antenna GUEs and $L$ single-antenna SUEs\footnote{Note that common types of SUEs include remote sensing satellites, spacecraft, space telescopes, space stations, planetary rovers and landers, and large-scale scientific experiment instruments. In Particular, unmanned aerial vehicles (UAVs) and high-altitude platforms (HAPs) can be regarded as special cases of SUEs that operate in super-near-ground environments.}, as shown in Fig. \ref{Fig1}. Specifically, BSs with MEC servers provide real-time computing services to GUEs, and LEO satellites with MEC servers supply the computing power for GUEs and SUEs\footnote{Here, we consider the new-generation LEO satellites with strong payload processing capabilities, which provide sufficient computing power resources for traditional GUEs and SUEs by deploying MEC servers to efficiently handle and analyze received data and conduct complex computing tasks.}. Each GUE and SUE has an indivisible computing task that needs to be offloaded to a MEC server for processing due to its own limited computing capability and stored energy. Note that the GUE is allowed to select at most one BS or LEO satellite for computing offloading, while the SUE adopts the concept of satellite clustering to select one LEO satellite from a group of N available options, i.e., $\sum\limits_{m = 1}^M {{\alpha _{k,m}} + } \sum\limits_{n = 1}^N {{\beta _{k,n}} = 1,} \forall k \in K$ and $\sum\limits_{n = 1}^N {{\gamma _{l,n}}}  = 1,\forall l \in L$, where binary variables ${\alpha _{k,m}}, {\beta _{k,n}}, {\gamma _{l,n}} \in \left\{ {0,1} \right\}$ indicate offloading selection for computing tasks. Particularly, if the task of the $k$-th GUE is offloaded to the $m$-th BS then ${\alpha _{k,m}} = 1$, if the task of the $k$-th GUE is offloaded to the $n$-th LEO satellite then ${\beta _{k,n}} = 1$, if the task of the $l$-th SUE is offloaded to the $n$-th LEO satellite then ${\gamma _{l,n}} = 1$, and otherwise the variable coefficient is equal to 0. In general, the whole computing offloading process consists of three stages. At first, GUEs and SUEs transmit the raw data of computing tasks to the BS or LEO satellite via uplink channels. Then, the BSs and LEO satellites decode the data of computing tasks, and perform the data computing at their own MEC servers. Finally, computing results are returned back to the corresponding GUEs and SUEs through downlink channels. Since the data size of computing results is much smaller than that of computing tasks, it is usual to ignore the transmission delay and energy consumption caused by the stage of returning the computing results \cite{computing results}. In the following, we describe the process of satellite-terrestrial computing in 6G wireless networks from the perspectives of communication and computing, respectively.

\subsection{Communication Model}
During the stage of data transmission, there involves three different channels\footnote{In this paper, we assume that the channel state information (CSI) of these three channels are completely available and unchanged during a time slot, but independently fade over time slots \cite{time slots}.}, i.e., inter-terrestrial channel, satellite-terrestrial channel and inter-satellite channel. Thus, we will discuss the communication models based on these three channels in turn.

\subsubsection{Terrestrial-to-terrestrial communication}
For the inter-terrestrial channel between the GUE and the BS, we consider both small-scale and large-scale fading. Specifically, the small-scale fading is assumed to obey Rayleigh model \cite{Rayleigh fading}, and the large-scale fading is modeled as a practical distance-dependent path loss attenuation, i.e., $\mathrm{PL}_{\mathrm{dB}}=128.1+37.6\log_{10}\tau$ \cite{Pass loss}, where $\tau$ (km) is the distance between the transmitter and receiver. To improve the spectrum utility, GUEs transmit data signals for computing offloading to the BSs in the NOMA manner. Then, successive interference cancellation (SIC) and receive beamforming technologies are adopted at the BS to suppress co-channel interference. In particular, we define ${\bf{h}}_{k,m}\in\mathbb{C}^{N_t^g\times1}$ is the channel vector from the $k$-th GUE to the $m$-th BS, and ${\pi ^m}\left( k \right)$ as the SIC order number of the $k$-th GUE at the $m$-th BS. If for any $k'$-th GUE and $k''$-th GUE with ${\left\| {{{\bf{h}}_{k',m}}} \right\|^2} > {\left\| {{{\bf{h}}_{k'',m}}} \right\|^2}$, then there is ${\pi ^m}\left( {k'} \right) < {\pi ^m}\left( {k''} \right)$, which represents that the signal of the $k'$-th GUE at the $m$-th BS is decoded and eliminated from the received signal before the $k''$-th GUE \cite{SIC}. Based on the principle of SIC that the GUE signal with a high channel quality is decoded first, the data transmission rate from the $k$-th GUE to the $m$-th BS is given by
\begin{equation}\label{GUE-BS rate}
R_{k,m}^{g - g}={B_1}{\log _2}\left( {1 + \frac{{{{\left| {{\bf{w}}_{k,m}^H{{\bf{h}}_{k,m}}} \right|}^2}{p_k}}}{{\sum\limits_{{{\pi ^m}\left( i \right) > {\pi ^m}\left( k \right)}} {{{\left| {{\bf{w}}_{k,m}^H{{\bf{h}}_{i,m}}} \right|}^2}{p_i} + \delta _1^2} }}} \right),
\end{equation}
where ${\bf{w}}_{k,m}\in\mathbb{C}^{N_t^g\times1}$ denotes a receive beamforming vector at the $m$-th BS with ${\bf{w}}_{k,m}^{H}{\bf{w}}_{k,m}=1$, $p_k$ is the transmit power of the $k$-th GUE, $B_1$ and $\delta _1^2$ represent the bandwidth and the variance of additive white Gaussian noise (AWGN) for the inter-terrestrial channel, respectively.

\subsubsection{Terrestrial-to-satellite communication}
According to the propagation characteristics of LEO satellite communication, the satellite-terrestrial channel between the $k$-th GUE and the $n$-th LEO satellite can be modeled as \cite{satellite-terrestrial channel} \cite{satellite-terrestrial channel2}
\begin{equation}\label{GUE-satellite channel}
{{\bf{g}}_{k,n}} = \sqrt {{C_{k,n}}} {\bf{b}}_{k,n}^{\frac{1}{2}}\odot{{\bf{r}}_{k,n}}\cdot\exp \left\{ {j2\pi {v_{k,n}^{{\rm{sat}}}}} \right\},
\end{equation}
where ${C_{k,n}}$, ${\bf{r}}_{k,n}$, ${{\bf{b}}_{k,n}}$ and $v_{k,n}^{{\rm{sat}}}$ denote the large-scale fading coefficient, the rain attenuation effect, the satellite antenna gain and Doppler shift caused by the motion of the $n$-th LEO satellite relative to the $k$-th GUE, respectively. In particular, ${C_{k,n}}$ is defined as
\begin{equation}\label{GUE-satellite large-scale fading}
{C_{k,n}} = {\left( {\frac{\mu }{{4\pi f{\varphi _{k,n}}}}} \right)^2}\frac{{{G_{k,n}}}}{{\kappa {B_2}T}},
\end{equation}
where $\mu$ is the speed of light, $f$ is the carrier frequency, $B_2$ is bandwidth of the satellite-terrestrial channel, $\kappa$ is Boltzmann constant, $T$ is the noise temperature, ${\varphi _{k,n}}$ and ${G_{k,n}}$ are the distance and transmit antenna gain from the $k$-th GUE to the $n$-th LEO satellite, respectively. In addition, the satellite-terrestrial channel is affected by various atmospheric attenuations in the troposphere, with rain attenuation having a dominant impact on channel quality. The rain attenuation vector can be expressed as \cite{rain attenuation}
\begin{equation}\label{rain attenuation}
{{\bf{r}}_{k,n}} = {\xi ^{\frac{1}{2}}}{e^{ - j{{\bm{\theta }}_{k,n}}}},
\end{equation}
where ${\xi ^{\frac{1}{2}}}$ denotes the power gain of the rain attenuation in dB, following log-normal distribution, i.e., $\ln \left( {{\xi ^{{1}/{2}}}} \right)\sim\mathcal{C}\mathcal{N}\left( {{\mu _r},\sigma _r^2} \right)$, and ${{\bm{\theta }}_{k,n}}$ is a phase vector whose components obey a uniform distributed between $0$ and $2\pi$. Moreover, the elements of the $N_t^s$-dimensional satellite receive antenna gain ${{\bf{b}}_{k,n}}$ are approximated by \cite{beam gain}
\begin{equation}\label{beam gain}
{{\bf{b}}_{k,n}}(i) = {b_{n,\max }}{\left( {\frac{{{J_1}\left( {{u_i}} \right)}}{{2{u_i}}} + 36\frac{{{J_3}\left( {{u_i}} \right)}}{{{u_i}^3}}} \right)^3},
\end{equation}
where ${u_i} = 2.07123\left( {{{\sin \left( {{\varepsilon _{i,k,n}}} \right)} \mathord{\left/{\vphantom {{\sin \left( {{\varepsilon _{i,k,n}}} \right)} {\sin \left( {\varepsilon _n^{3dB}} \right)}}} \right.\kern-\nulldelimiterspace} {\sin \left( {\varepsilon _n^{3dB}} \right)}}} \right)$ with ${\varepsilon _{i,k,n}}$ being the angle between the $i$-th antenna of the $n$-th LEO satellite and the $k$-th GUE, and the constant $\varepsilon _n^{3dB}$ being 3-dB angle for the $n$-th LEO satellite, ${b_{n,\max }}$ is the maximum antenna gain of the $n$-th LEO satellite, ${J_1}$ and ${J_3}$ are the first-kind Bessel functions of the first and third order, respectively. Similarly, GUEs transmit signals for computing tasks to the LEO satellites by NOMA over the uplink satellite-terrestrial channel. Then, LEO satellites perform Doppler compensation on the received signals and apply receive beamforming to decode the data from GUEs by SIC.\footnote{In satellite-terrestrial networks, the high speed mobility of satellites relative to the ground can cause Doppler shift effects, which have a serious impact on communication performance. Doppler compensation is therefore required at the satellite receivers to improve signal transmission efficiency and accuracy \cite{Doppler}.}
Similarly, we define ${\pi ^n}\left( k \right)$ as the SIC order number of the $k$-th GUE at the $n$-th LEO satellite. If for any $k'$-th GUE and $k''$-th GUE with ${\left\| {{{\bf{g}}_{k',n}}} \right\|^2} > {\left\| {{{\bf{g}}_{k'',n}}} \right\|^2}$, then there is ${\pi ^n}\left( {k'} \right) < {\pi ^n}\left( {k''} \right)$, which represents that the signal of the $k'$-th GUE at the $n$-th LEO satellite is decoded and eliminated from the received signal before the $k''$-th GUE. In this context, the data transmission rate from the $k$-th GUE to the $n$-th LEO satellite is given by
\begin{equation}\label{GUE-Satellite rate}
R_{k,n}^{g - s} = {B_2}{\log _2}\left( {1 + \frac{{{{\left| {{\bf{v}}_{k,n}^H{{\bf{g}}_{k,n}}} \right|}^2}{p_k}}}{{\sum\limits_{{{\pi ^n}\left( i \right) > {\pi ^n}\left( k \right)}} {{{\left| {{\bf{v}}_{k,n}^H{{\bf{g}}_{i,n}}} \right|}^2}{p_i} + \delta _2^2} }}} \right),
\end{equation}
where ${\bf{v}}_{k,n}\in\mathbb{C}^{N_t^s\times1}$ denotes receive beamforming at the $n$-th LEO satellite for the $k$-th GUE with ${\bf{v}}_{k,n}^{H}{\bf{v}}_{k,n}=1$, and $\delta _2^2$ denotes the variance of AWGN for the satellite-terrestrial channel.

\subsubsection{Satellite-to-satellite communication}
For the inter-satellite channel between the SUE and the LEO satellite, free space optical (FSO) communication is used to reduce inter-user interference and improve communication security for achieving the long-range and high-speed data transmission. According to the characteristics of FSO communication described in \cite{FSO 1} and \cite{FSO 2}, the data transmission rate from the $l$-th SUE to the $n$-th LEO satellite can be expressed as
\begin{equation}\label{SUE-Satellite rate}
R_{l,n}^{s - s} = {B_3}{\log _2}\left( {1 + \frac{{{q_l}\eta _l^t\eta _n^r{{\left( {\frac{\lambda }{{4\pi {\phi _{l,n}}}}} \right)}^2}G_l^tG_n^rL_{l}^tL_{n}^r}}{{\delta _3^2}}} \right),
\end{equation}
where $q_l$ is the transmit power of the $l$-th SUE, $\lambda $ is the wavelength, ${\phi _{l,n}}$ is the distance between the $l$-th SUE and the $n$-th LEO satellite, ${B_3}$ and $\delta _3^2$ represent the bandwidth and the variance of AWGN for the inter-satellite channel, respectively. $G_l^t = {( {\pi D_l^t/\lambda } )^2}$, $L_{l}^t = \exp ( { - G_l^t{{( {e_{l}^t} )}^2}} )$, $D_l^t$, $e_{l}^t$ and $\eta _l^t$ are the aperture gain, the pointing loss factor, the aperture diameter, the pointing error angle and the optical efficiency of transmitter at the $l$-th SUE, respectively. Similarly, $G_n^r = {( {\pi D_n^r/\lambda } )^2}$, $L_{n}^r = \exp ( { - G_n^r{{( {e_{n}^r} )}^2}} )$, $D_n^r$, $e_{n}^r$ and $\eta _n^r$ are corresponding parameters of receiver at the $n$-th LEO satellite.

\subsection{Computing Model}
In general, execution time and energy consumption are two key performance metrics for completing computing tasks. Thus, we characterize the computing model from these two aspects. For easy notation, the computing task of the $k$-th GUE is defined as $\Omega _k^g \triangleq \left( {{d_k},{c_k}} \right)$, where ${d_k}$ denotes the input data size in bits and ${c_k}$ denotes task complexity in cycles/bit which means the number of CPU cycles required to compute per bit of input data. Similarly, the computing task of the $l$-th SUE is defined as $\Omega _l^s \triangleq \left( {d_l^{space},c_l^{space}} \right)$, where $d_l^{space}$ and $c_l^{space}$ are the input data size and task complexity of the SUE task, respectively. In what follows, we analyze the performance of execution time and energy consumption for completing computing tasks from three different computing models, respectively.

\subsubsection{Terrestrial-to-terrestrial computing}
According to the inter-terrestrial communication model in (\ref{GUE-BS rate}), the uplink transmission delay for computing task $\Omega _k^g$ from the $k$-th GUE to the BS can be computed as
\begin{equation}\label{T_k^g-g,tra}
T_k^{g - g,tra} = \sum\limits_{m = 1}^M {{\alpha _{k,m}}\frac{{{d_k}}}{{R_{k,m}^{g - g}}}}.
\end{equation}
Then, the decoded computing task $\Omega_k^g$ is sent to the MEC server by the BS for data computing. The computing delay at the MEC server is given by
\begin{equation}\label{T_k^{g - g,com}}
T_k^{g - g,com} = \sum\limits_{m = 1}^M {{\alpha _{k,m}}\frac{{{d_k}{c_k}}}{{f_{k,m}^{gro}}}},
\end{equation}
where ${f_{k,m}^{gro}}\geq0$ denotes the computing resources allocated to the $k$-th GUE by the MEC server of the $m$-th BS. Therefore, the total execution time for completing the computing task of the $k$-th GUE can be expressed as
\begin{equation}\label{T_k^{g - g}}
T_k^{g - g} = T_k^{g - g,tra} + T_k^{g - g,com}.
\end{equation}
Moreover, the transmission energy consumption from the $k$-th GUE to the BS is given by
\begin{equation}\label{E_k^{g - g,tra}}
E_k^{g - g,tra} = \sum\limits_{m = 1}^M {{\alpha _{k,m}}{p_k}\frac{{{d_k}}}{{R_{k,m}^{g - g}}}}.
\end{equation}
The computing energy consumption for processing the task can be modeled as \cite{computing energy consumption}
\begin{equation}\label{E_k^{g - g,com}}
E_k^{g - g,com} = \sum\limits_{m = 1}^M {{\alpha _{k,m}}\tau _m^{gro}{d_k}{c_k}{{\left( {f_{k,m}^{gro}} \right)}^2}},
\end{equation}
where $\tau _m^{gro}$ is the energy coefficient of the MEC server at the $m$-th BS, which is  related to the chip architecture \cite{energy coefficient}. Hence, the total energy consumption for computing offloading from the $k$-th GUE to the BS can be expressed as
\begin{equation}\label{E_k^{g - g}}
E_k^{g - g} = E_k^{g - g,tra} + E_k^{g - g,com}.
\end{equation}

\subsubsection{Terrestrial-to-satellite computing}
Compared with terrestrial-to-terrestrial computing, terrestrial-to-satellite computing with very long transmission distance needs to consider the propagation delay \cite{satellite network}. Thus, the total execution time consists of three components, i.e., transmission delay $T_k^{g - s,tra}$, propagation delay $T_k^{g - s,pro}$ and computing delay $T_k^{g - s,com}$, which can be expressed as
\begin{align}\label{T_k^{g - s}}
&T_k^{g - s} = T_k^{g - s,tra} + T_k^{g - s,pro} + T_k^{g - s,com}\notag\\
&\ =\sum\limits_{n = 1}^N {{\beta _{k,n}}\frac{{{d_k}}}{{R_{k,n}^{g - s}}}}+\sum\limits_{n = 1}^N {{\beta _{k,n}}\frac{{{\varphi _{k,n}}}}{\mu }}+\sum\limits_{n = 1}^N {{\beta _{k,n}}\frac{{{d_k}{c_k}}}{{f_{k,n}^{sat-g}}}},
\end{align}
where ${f_{k,n}^{sat - g}}\geq0$ denotes the computing resources allocated to the $k$-th GUE by the MEC server of the $n$-th LEO satellite.
Likewise, the total energy consumption for computing offloading from the $k$-th GUE to the LEO satellite can be expressed as
\begin{align}\label{E_k^{g - s}}
&E_k^{g - s} = E_k^{g - s,tra} + E_k^{g - s,com}\notag\\
&\ \ =\sum\limits_{n = 1}^N {{\beta _{k,n}}{p_k}\frac{{{d_k}}}{{R_{k,n}^{g - s}}}}+\sum\limits_{n = 1}^N {{\beta _{k,n}}\tau _n^{sat}{d_k}{c_k}{{\left( {f_{k,n}^{sat-g}} \right)}^2}},
\end{align}
where $\tau _m^{sat}$ is the energy coefficient of the MEC server at the $n$-th LEO satellite, which is determined by the chip architecture.

\subsubsection{Satellite-to-satellite computing}
Similar to terrestrial-to-satellite computing, the total execution time for completing the computing tasks of SUEs also includes transmission delay, propagation delay and computing delay, namely
\begin{align}\label{T_l^{s - s}}
&T_l^{s - s} = T_l^{s - s,tra} + T_l^{s - s,pro} + T_l^{s - s,com}\notag\\
&\ \ =\sum\limits_{n = 1}^N {{\gamma _{l,n}}\frac{{d_l^{space}}}{{R_{l,n}^{s - s}}}}+\sum\limits_{n = 1}^N {{\gamma _{l,n}}\frac{{{\phi _{l,n}}}}{\mu }}+\sum\limits_{n = 1}^N {{\gamma _{l,n}}\frac{{d_l^{space}c_l^{space}}}{{f_{l,n}^{sat - s}}}},
\end{align}
where ${f_{l,n}^{sat - s}}\geq0$ denotes the computing resources allocated to the $l$-th SUE by the MEC server of the $n$-th LEO satellite.
Correspondingly, the total energy consumption for computing offloading from the $l$-th SUE to the LEO satellite is given by
\begin{align}\label{E_l^{s - s}}
&E_l^{s - s} = E_l^{s - s,tra} + E_l^{s - s,com}\notag\\
&\ \ =\sum\limits_{n = 1}^N {{\gamma _{l,n}}{q_l}\frac{{d_l^{space}}}{{R_{l,n}^{s - s}}}}+\sum\limits_{n = 1}^N {{\gamma _{l,n}}\tau _n^{sat}d_l^{space}c_l^{space}{{\left( {f_{l,n}^{sat - s}} \right)}^2}}.
\end{align}

It is observed that the offloading selection $\{{\alpha _{k,m}}, {\beta _{k,n}}, {\gamma _{l,n}}\}$, beamforming design $\{{\bf{w}}_{k,m}, {\bf{v}}_{k,n}\}$, and resource allocation $\{p_k, q_l, f_{k,m}^{gro}, f_{k,n}^{sat - g}, f_{l,n}^{sat - s}\}$ all have important effects on the execution time and energy consumption for completing the computing tasks. Therefore, it makes sense to develop a joint design to reduce the time and energy costs for satellite-terrestrial computing in 6G wireless networks.

\section{Joint Design for Satellite-Terrestrial Computing}
In this section, we provide an energy-efficient design for satellite-terrestrial computing by jointly optimizing offloading selection, beamforming design and resource allocation.

\subsection{Problem Formulation}
Considering that user experience is mainly determined by the energy consumption and the execution time of completing tasks, we aim to minimize the weighted total energy consumption while ensuring delay requirements of computing tasks. In this context, the joint design can be mathematically formulated as the following optimization problem:
\begin{subequations}\label{OP0}
\begin{align}
\mathop {\min }\limits_{\bm{\alpha} ,\bm{\beta} ,\bm{\gamma} ,\bm{w},\bm{v},\bf{p},\bf{q},\bf{f}} & \sum\limits_{k = 1}^K {\rho _k^g\left( {E_k^{g - g} + E_k^{g - s}} \right)}  + \sum\limits_{l = 1}^L {\rho _l^sE_l^{s - s}} {\rm{ }}\label{OP0obj}\\
\textrm{s.t.}\ \ \
&T_k^{g - g} + T_k^{g - s} \le Z_k^g,\label{OP0st1}\\
&T_l^{s - s} \le Z_l^s,\label{OP0st2}\\
&\sum\limits_{k = 1}^K {{\alpha _{k,m}}f_{k,m}^{gro}}  \le F_m^{gro},\label{OP0st3}\\
&\sum\limits_{k = 1}^K {{\beta _{k,n}}f_{k,n}^{sat - g}}  + \sum\limits_{l = 1}^L {{\gamma _{l,n}}f_{l,n}^{sat - s}}  \le F_n^{sat},\label{OP0st4}\\
&{\alpha _{k,m}}, {\beta _{k,n}}, {\gamma _{l,n}} \in \left\{ {0,1} \right\},\label{OP0st5}\\
&\sum\limits_{m = 1}^M {{\alpha _{k,m}} + } \sum\limits_{n = 1}^N {{\beta _{k,n}} = 1},\label{OP0st6}\\
&\sum\limits_{n = 1}^N {{\gamma _{l,n}}}  = 1,\label{OP0st7}\\
&0 \le {p_k} \le P_k^{\max },\label{OP0st8}\\
&0 \le {q_l} \le Q_l^{\max },\label{OP0st9}\\
& \|{\bf{w}}_{k,m}\|^2 = \|{\bf{v}}_{k,n}\|^2 = 1,\label{OP0st10}
\end{align}
\end{subequations}
where $\bm{\alpha} = \{{\alpha _{k,m}}, \forall k \in K, \forall m \in M\}$, $\bm{\beta} = \{{\beta _{k,n}}, \forall k \in K, \forall n \in N\}$, $\bm{\gamma} = \{{\gamma _{l,n}}, \forall l \in L, \forall n \in N\}$, ${\bm{w}} = \{{\bf{w}}_{k,m}, \forall k \in K, \forall m \in M\}$, ${\bm{v}} = \{{\bf{v}}_{k,n}, \forall k \in K, \forall n \in N\}$, ${\bf{p}} = \{{p_k}, \forall k \in K\}$, ${\bf{q}} = \{{q_l}, \forall l \in L\}$ and ${\bf{f}} = \{f_{k,m}^{gro}, \forall k \in K, \forall m \in M \} \bigcup \{f_{k,n}^{sat - g}, \forall k \in K, \forall n \in N \} \bigcup \{f_{l,n}^{sat - s}, \forall l \in L, \forall n \in N \}$ are the collections of optimization variables. For problem (\ref{OP0}), the objective function (\ref{OP0obj}) is the weighted sum of energy consumption, where $\rho _k^g$ and $\rho _l^s$ are the energy weights of the $k$-th GUE and the $l$-th SUE, respectively. Constraints (\ref{OP0st1}) and (\ref{OP0st2}) are the execution time requirements with $Z_k^g$ and $Z_l^s$ being the maximum delay of the $k$-th GUE and the $l$-th SUE, respectively. Constrains (\ref{OP0st3})-(\ref{OP0st4}) describe computing resources restrictions imposed by MEC servers at the BSs and the LEO satellites with $F_m^{gro}$ and $F_n^{sat}$ being the maximum computing power of the MEC server at the $m$-th BS and the $n$-th LEO satellite, respectively. Constrains (\ref{OP0st5})-(\ref{OP0st7}) are the rules of offloading selection, namely the GUE is allowed to select at most one BS or LEO satellite for computing offloading, while the SUE can only select at most one LEO satellite. Constraints (\ref{OP0st8}) and (\ref{OP0st9}) mean transmit power limitations, where $P_k^{\max}$ and $Q_l^{\max}$ are the maximum transmit power budget of the $k$-th GUE and the $l$-th SUE, respectively. Finally, constraint (\ref{OP0st10}) denotes the normalized receive beamforming at the BSs and the LEO satellites. Note that problem (\ref{OP0}) is a typical mixed-integer nonlinear programming (MINLP) problem, which is proved to be NP-hard and is extremely intractable to obtain its optimal solution in polynomial time \cite{nphard}. To this end, we turn to develop an effective algorithm for finding a feasible sub-optimal solution to achieve a competitive performance of satellite-terrestrial computing in 6G wireless networks.

\subsection{Algorithm Design}
It is seen that the optimization variables are coupled in the objective function and the constraints, causing problem (\ref{OP0}) to be unmanageable. To address this challenge, we adopt the commonly used alternating optimization (AO) method \cite{AO} to decompose problem (\ref{OP0}) into three subproblems, i.e., offloading selection subproblem, beamforming design subproblem and resource allocation subproblem. Now, we consider the first subproblem that optimizing offloading selection while fixing others, which is given by
\begin{subequations}\label{OP1}
\begin{align}
\mathop {\min }\limits_{\bm{\alpha} ,\bm{\beta} ,\bm{\gamma}} & \sum\limits_{k = 1}^K {\rho _k^g\left( {E_k^{g - g} + E_k^{g - s}} \right)}  + \sum\limits_{l = 1}^L {\rho _l^sE_l^{s - s}} {\rm{ }}\label{OP1obj}\tag{19}\\
\textrm{s.t.}\
&(\ref{OP0st1})-(\ref{OP0st7})\notag.
\end{align}
\end{subequations}
Notice that problem (\ref{OP1}) is a binary integer programming problem, which can be solved by using the well-known branch and bound (B\&B) method \cite{BnB}. Specifically, problem (\ref{OP1}) can be considered as the root node of the B\&B search tree, which is constructed and traversed by the processes of branching and bounding. In short, branching is to divide a parent problem into two subproblems by adding binary constraints on the nodes, and bounding is to check the upper and lower bounds of the subproblems during branching. In this context, we need to find the lower bound of problem (\ref{OP1}) by solving the following relaxed optimization:
\begin{subequations}\label{OP1-1}
\begin{align}
\mathop {\min }\limits_{\bm{\alpha} ,\bm{\beta} ,\bm{\gamma}} & \sum\limits_{k = 1}^K {\rho _k^g\left( {E_k^{g - g} + E_k^{g - s}} \right)}  + \sum\limits_{l = 1}^L {\rho _l^sE_l^{s - s}} {\rm{ }}\label{OP1-1obj}\\
\textrm{s.t.}\
&(\ref{OP0st1})-(\ref{OP0st4}),(\ref{OP0st6}),(\ref{OP0st7})\notag,\\
&{\alpha _{k,m}}, {\beta _{k,n}}, {\gamma _{l,n}} \in \left[ {0, 1} \right],\label{OP1-1st1}
\end{align}
\end{subequations}
where binary variables $\bm{\alpha}, \bm{\beta}$ and $\bm{\gamma}$ are relaxed to be continuous with a range of 0 to 1. In this case, problem (\ref{OP1-1}) is a convex optimization problem, whose optimal solution is just the lower bound of the objective value of problem (\ref{OP1}). Thus, the optimal solution of problem (\ref{OP1}) can be found by solving problem (\ref{OP1-1}) with the B\&B algorithm during the operations of branching and bounding. Although the B\&B algorithm can provide superior and optimal solutions for small and medium-sized problems, it is not an effective method for handling large-scale user offloading selection cases due to its exponential search space in the worst case. Therefore, for the offloading selection subproblem, we adopt a relaxation mapping method with lower complexity. Specifically, after the 0-1 constraint ${\alpha _{k,m}}, {\beta _{k,n}}, {\gamma _{l,n}} \in \left\{ {0,1}\right\}$ is relaxed to ${\alpha _{k,m}}, {\beta _{k,n}}, {\gamma _{l,n}} \in \left[ {0, 1} \right]$, the practical significance of the optimization problem (\ref{OP1-1}) can be regarded as each computing task can be partitioned into multiple parts to complete the computing offloading, so that ${\alpha _{k,m}}$ and ${\beta _{k,n}}$ denote the offloading portion of the $k$-th GUE's computing task for the $m$-th BS and the $n$-th LEO satellite, respectively, and ${\gamma _{l,n}}$ denotes the offloading portion of the $l$-th SUE's computing task for the $n$-th LEO satellite. However, the optimal solution to the standard linear optimization problem (\ref{OP1-1}) is continuous and needs to be mapped to 0-1 variables. The specific mapping strategy is to compare the values of ${\alpha _{k,m}},\forall m\in M$ and ${\beta _{k,n}},\forall n\in N$ for the $k$-th GUE, mapping the largest of them to 1 and the others to 0, and compare the values of ${\gamma _{l,n}},\forall n\in N$ for the $l$-th SUE, mapping the largest one to 1 and the others to 0. In this way, we obtain a suboptimal solution to the offloading selection subproblem (\ref{OP1}), and since problem (\ref{OP1}) is a standard linear discrete programming, the suboptimal solution obtained through the aforementioned relaxation mapping method approximates the optimal solution \cite{related work 5}.

Next, we consider the beamforming design subproblem, which is formulated as
\begin{subequations}\label{OP2}
\begin{align}
\mathop {\min }\limits_{\bm{w},\bm{v}} & \sum\limits_{k = 1}^K {\rho _k^g\left( {E_k^{g - g,tra} + E_k^{g - s,tra}} \right)}\label{OP2obj}\tag{21}\\
\textrm{s.t.}\
&(\ref{OP0st1}),(\ref{OP0st10}),\notag\
\end{align}
\end{subequations}
where constraints (\ref{OP0st2})-(\ref{OP0st9}) that do not involve beamforming vectors $\bm{w}$ and $\bm{v}$ are not considered.
Note that problem (\ref{OP2}) is nonconvex because the variables $\bm{w}$ and $\bm{v}$ exist in quadratic fractional form in the objective function and constraint (\ref{OP0st1}). To solve this issue, we introduce the auxiliary variables $A_{k,m}^g$ and $A_{k,n}^s$, and then transform the original problem (\ref{OP2}) as
\begin{subequations}\label{OP2-1}
\begin{align}
&\mathop {\min }\limits_{\bm{w},\bm{v},A_{k,m}^g,A_{k,n}^s} \sum\limits_{k = 1}^K {\rho _k^g\left({\sum\limits_{m = 1}^M {{\alpha _{k,m}}{p_k}A_{k,m}^g}  + \sum\limits_{n = 1}^N {{\beta _{k,n}}{p_k}A_{k,n}^s} }\right)} \label{OP2-1obj}\\
&\ \ \ \ \textrm{s.t.}\
(\ref{OP0st10}),\notag\\
&\ \ \ \ \ \ \ \frac{{{d_k}}}{{R_{k,m}^{g - g}}} \le A_{k,m}^g,\label{OP2-1st1}\\
&\ \ \ \ \ \ \ \frac{{{d_k}}}{{R_{k,n}^{g - s}}} \le A_{k,n}^s,\label{OP2-1st2}\\
&\ \ \ \ \ \ \ \sum\limits_{m = 1}^M {{\alpha _{k,m}}A_{k,m}^g}  + \sum\limits_{n = 1}^N {{\beta _{k,n}}A_{k,n}^s}\le Z_k^g - \sum\limits_{m = 1}^M {\alpha _{k,m}}\frac{{{d_k}{c_k}}}{{f_{k,m}^{gro}}}\notag\\
&\ \ \ \ \ \ \ \ \ \ \ \ \ \ \ - \sum\limits_{n = 1}^N {{\beta _{k,n}}\frac{{{d_k}{c_k}}}{{f_{k,n}^{sat - g}}}}  - \sum\limits_{n = 1}^N {{\beta _{k,n}}\frac{{{\varphi _{k,n}}}}{\mu }},\label{OP2-1st3}
\end{align}
\end{subequations}
where constraint (\ref{OP0st1}) is substituted for constraints (\ref{OP2-1st1})-(\ref{OP2-1st3}), but constraints (\ref{OP2-1st1}) and (\ref{OP2-1st2}) are still noncovex. To address the noncovexity of constraint (\ref{OP2-1st1}), we bring in new auxiliary variables $\tilde R_{k,m}^{g - g}$ and $\tilde \Gamma _{k,m}^{g - g}$. In this context, (\ref{OP2-1st1}) can be replaced by
\begin{equation}\label{OP2-1st1-1}
\frac{{{d_k}}}{{\tilde R_{k,m}^{g - g}}} \le A_{k,m}^g,
\end{equation}
\begin{equation}\label{OP2-1st1-2}
\tilde R_{k,m}^{g - g} \le {B_1}{\log _2}\left( {1 + \tilde \Gamma _{k,m}^{g - g}} \right),
\end{equation}
and
\begin{align}\label{OP2-1st1-3}
&\sum\limits_{{{\pi ^m}\left( i \right) > {\pi ^m}\left( k \right)}} {\text{tr}({\mathbf{h}_{i,m}}\mathbf{h}_{i,m}^H{\mathbf{w}_{k,m}}\mathbf{w}_{k,m}^H){p_i} + \delta _1^2}\notag \\
&\ \ \ \ \ \le \frac{{\text{tr}({\mathbf{h}_{k,m}}\mathbf{h}_{k,m}^H{\mathbf{w}_{k,m}}\mathbf{w}_{k,m}^H){p_k}}}{{\tilde \Gamma _{k,m}^{g - g}}}.
\end{align}
Similarly, through introducing auxiliary variables $\tilde R_{k,n}^{g - s}$ and $\tilde \Gamma _{k,n}^{g - s}$, nonconvex constraint (\ref{OP2-1st2}) can be converted as
\begin{equation}\label{OP2-1st2-1}
\frac{{{d_k}}}{{\tilde R_{k,n}^{g - s}}} \le A_{k,n}^s,
\end{equation}
\begin{equation}\label{OP2-1st2-2}
\tilde R_{k,n}^{g - s} \le {B_2}{\log _2}\left( {1 + \tilde \Gamma _{k,n}^{g - s}} \right),
\end{equation}
and
\begin{align}\label{OP2-1st2-3}
&\sum\limits_{{{\pi ^n}\left( i \right) > {\pi ^n}\left( k \right)}} {\text{tr}({\mathbf{g}_{i,n}}\mathbf{g}_{i,n}^H{\mathbf{v}_{k,n}}\mathbf{v}_{k,n}^H){p_i} + \delta _2^2}  \notag \\
&\ \ \ \ \ \le \frac{{\text{tr}({\mathbf{g}_{k,n}}\mathbf{g}_{k,n}^H{\mathbf{v}_{k,n}}\mathbf{v}_{k,n}^H){p_k}}}{{\tilde \Gamma _{k,n}^{g - s}}}.
\end{align}
However, additional nonconvex constraints (\ref{OP2-1st1-3}) and (\ref{OP2-1st2-3}) block the solve of the problem. To this end, we utilize the successive convex approximation (SCA) technique to handle them. In particular, the binary first-order Taylor series expansion is applied in the right-hand part of the inequality (\ref{OP2-1st1-3}) at point $( {\mathbf{w}_{k,m}^\# ,{\tilde \Gamma }_{k,m}^{\#g - g}} )$, where $\mathbf{w}_{k,m}^\# $ and ${\tilde \Gamma}_{k,m}^{\#g - g}$ are the value of ${\mathbf{w}_{k,m}}$ and $\tilde \Gamma _{k,m}^{g - g}$ in the last iteration, respectively. Hence, constraint (\ref{OP2-1st1-3}) can be rewritten as
\begin{align}\label{Taylor1}
&\sum\limits_{{{\pi ^m}\left( i \right) > {\pi ^m}\left( k \right)}} {\text{tr}({\mathbf{h}_{i,m}}\mathbf{h}_{i,m}^H{\mathbf{w}_{k,m}}\mathbf{w}_{k,m}^H){p_i} + \delta _1^2}  \notag \\
&\le \left( {\frac{{{X_{k,m}}{{\tilde \Gamma }}_{k,m}^{\# g - g} - X_{k,m}^\# \tilde \Gamma _{k,m}^{g - g} + X_{k,m}^\# {{\tilde \Gamma }}_{k,m}^{\# g - g}}}{{{{\left( {{{\tilde \Gamma }}_{k,m}^{\# g - g}} \right)}^2}}}} \right){p_k},
\end{align}
where ${X_{k,m}} = \text{tr}\left({\mathbf{h}_{k,m}}\mathbf{h}_{k,m}^H{\mathbf{w}_{k,m}}\mathbf{w}_{k,m}^H\right)$ and $X_{k,m}^\#  = \text{tr}\left({\mathbf{h}_{k,m}}\mathbf{h}_{k,m}^H\mathbf{w}_{k,m}^\# \mathbf{w}{_{k,m}^\#}{}^H\right)$. Likewise, constraint (\ref{OP2-1st2-3}) can be rewritten as
\begin{align}\label{Taylor2}
&\sum\limits_{{{\pi ^n}\left( i \right) > {\pi ^n}\left( k \right)}} {\text{tr}({\mathbf{g}_{i,n}}\mathbf{g}_{i,n}^H{\mathbf{v}_{k,n}}\mathbf{v}_{k,n}^H){p_i} + \delta _2^2}  \notag \\
&\le \left( {\frac{{{Y_{k,n}}{{\tilde \Gamma }}_{k,n}^{\# g - s} - Y_{k,n}^\# \tilde \Gamma _{k,n}^{g - s} + Y_{k,n}^\# {{\tilde \Gamma }}_{k,n}^{\# g - s}}}{{{{\left( {{{\tilde \Gamma }}_{k,n}^{\# g - s}} \right)}^2}}}} \right){p_k},
\end{align}
where ${Y_{k,n}} = \text{tr}\left({\mathbf{g}_{k,n}}\mathbf{g}_{k,n}^H{\mathbf{v}_{k,n}}\mathbf{v}_{k,n}^H\right)$ and $Y_{k,n}^\#  = \text{tr}\left({\mathbf{g}_{k,n}}\mathbf{g}_{k,n}^H\mathbf{v}_{k,n}^\# \mathbf{v}{_{k,n}^\#}{}^H\right)$. $\mathbf{v}_{k,n}^\#$ and ${{\tilde \Gamma }}_{k,n}^{\# g - s}$ are the values of $\mathbf{v}_{k,n}$ and ${{\tilde \Gamma }}_{k,n}^{g - s}$ in the last iteration, respectively. Furthermore, by using the semi-definite relaxation (SDR) technique, i.e., ${\mathbf{W}_{k,m}} = {\mathbf{w}_{k,m}}\mathbf{w}_{k,m}^H$ and ${\mathbf{V}_{k,n}} = {\mathbf{v}_{k,n}}\mathbf{v}_{k,n}^H$, the beamforming design subproblem (\ref{OP2-1}) can be reconstructed as
\begin{subequations}\label{OP2-2}
\begin{align}
&\mathop {\min }\limits_{\Lambda} \sum\limits_{k = 1}^K {\rho _k^g\left({\sum\limits_{m = 1}^M {{\alpha _{k,m}}{p_k}A_{k,m}^g}  + \sum\limits_{n = 1}^N {{\beta _{k,n}}{p_k}A_{k,n}^s} }\right)}\label{OP2-2obj}\\
&\textrm{s.t.}\
(\ref{OP2-1st3}),(\ref{OP2-1st1-1}),(\ref{OP2-1st1-2}),(\ref{OP2-1st2-1}),(\ref{OP2-1st2-2}),\notag\\
&\sum\limits_{{{\pi ^m}\left( i \right) > {\pi ^m}\left( k \right)}} {\text{tr}({\mathbf{h}_{i,m}}\mathbf{h}_{i,m}^H{\mathbf{W}_{k,m}}){p_i} + \delta _1^2}  \notag\\
&\le \left( {\frac{{{X_{k,m}}{{\tilde \Gamma }}_{k,m}^{\# g - g} - X_{k,m}^\# \tilde \Gamma _{k,m}^{g - g} + X_{k,m}^\# {{\tilde \Gamma }}_{k,m}^{\# g - g}}}{{{{\left( {{{\tilde \Gamma }}_{k,m}^{\# g - g}} \right)}^2}}}} \right){p_k},\label{OP2-2st1}\\
&\sum\limits_{{{\pi ^n}\left( i \right) > {\pi ^n}\left( k \right)}} {\text{tr}({\mathbf{g}_{i,n}}\mathbf{g}_{i,n}^H{\mathbf{V}_{k,n}}){p_i} + \delta _2^2}  \notag\\
&\ \le \left( {\frac{{{Y_{k,n}}{{\tilde \Gamma }}_{k,n}^{\# g - s} - Y_{k,n}^\# \tilde \Gamma _{k,n}^{g - s} + Y_{k,n}^\# {{\tilde \Gamma }}_{k,n}^{\# g - s}}}{{{{\left( {{{\tilde \Gamma }}_{k,n}^{\# g - s}} \right)}^2}}}} \right){p_k},\label{OP2-2st2}\\
&\ \ \ {\mathbf{W}_{k,m}} \succeq 0,\ {\mathbf{V}_{k,n}} \succeq 0,\label{OP2-2st3}\\
&\ \ \ \text{tr}(\mathbf{W}_{k,m})=\text{tr}(\mathbf{V}_{k,n})=1,\label{OP2-2st4}
\end{align}
\end{subequations}
where $\Lambda \triangleq \{\mathbf{W}_{k,m}, \mathbf{V}_{k,n}, A_{k,m}^g, A_{k,n}^s, \tilde R_{k,m}^{g - g}, \tilde \Gamma _{k,m}^{g - g}, \tilde R_{k,n}^{g - s},$ $\tilde \Gamma _{k,n}^{g - s}, \forall k \in K, \forall m \in M, \forall n \in N\}$. It is worth pointing that the nonconvex rank-one constraints $\text{Rank}({\mathbf{W}_{k,m}}) = 1$ and $\text{Rank}({\mathbf{V}_{k,n}}) = 1$ in problem (\ref{OP2-2}) have been dropped because it is proved in Appendix A that the solutions always meet the rank-one constraints. Through a series of transformation, problem (\ref{OP2-2}) become a standard convex optimization problem, which can be directly solved by some optimization toolboxes, such as CVX \cite{CVX}. As a result, after obtaining the optimal solution $({\mathbf{W}_{k,m}^{*}},{\mathbf{V}_{k,n}^{*}})$ to problem (\ref{OP2-2}), the optimal solution $({\mathbf{w}_{k,m}^{*}},{\mathbf{v}_{k,n}^{*}})$ to problem (\ref{OP2}) can be obtained via eigenvalue decomposition (EVD).

Finally, the resource allocation subproblem by jointly optimizing transmit power and computing power can be formulated as
\begin{subequations}\label{OP3}
\begin{align}
\mathop {\min }\limits_{\bf{p},\bf{q},\bf{f}} & \sum\limits_{k = 1}^K {\rho _k^g\left( {E_k^{g - g} + E_k^{g - s}} \right)}  + \sum\limits_{l = 1}^L {\rho _l^sE_l^{s - s}}\label{OP3obj}\tag{32}\\
\textrm{s.t.}\
&(\ref{OP0st1})-(\ref{OP0st4}),(\ref{OP0st8}),(\ref{OP0st9}).\notag
\end{align}
\end{subequations}
Since each GUE is allowed to select at most one BS or LEO satellite for computing offloading, problem (\ref{OP3}) can be solved by two ways in terms of the transmit power of GUEs.
Specifically, if $\alpha_{k,m}=1$, the transmission energy consumption from the $k$-th GUE to the BS can be expressed as a function of the transmit power $p_k$, i.e.,
\begin{equation}\label{E(p)1}
E_k^{g - g,tra}\left( {{p_k}} \right) = \sum\limits_{m = 1}^M {{}{p_k}\frac{{{d_k}}}{{{B_1}{{\log }_2}\left( {1 + \frac{{{{| {\mathbf{w}_{k,m}^H{\mathbf{h}_{k,m}}} |}^2}{p_k}}}{{ {I^g_{k,m} + \delta _1^2} }}} \right)}}},
\end{equation}
where $I^g_{k,m}=\sum\limits_{{{\pi ^m}\left( i \right)>{\pi ^m}\left( k \right)}} {{| {\mathbf{w}_{k,m}^H{\mathbf{h}_{i,m}}} |}^2}{p_i}$. And if $\beta_{k,n}=1$, the transmission energy consumption from the $k$-th GUE to the LEO satellite is also a function of $p_k$, which is given by
\begin{equation}\label{E(p)2}
E_k^{g - s,tra}\left( {{p_k}} \right) = \sum\limits_{n = 1}^N {{}{p_k}\frac{{{d_k}}}{{{B_2}{{\log }_2}\left( {1 + \frac{{{{| {\mathbf{v}_{k,n}^H{\mathbf{g}_{k,n}}} |}^2}{p_k}}}{{I^s_{k,n} + \delta _2^2} }} \right)}}},
\end{equation}
where $I^s_{k,n}=\sum\limits_{{{\pi ^n}\left( i \right)>{\pi ^n}\left( k \right)}} {{| {\mathbf{v}_{k,n}^H{\mathbf{g}_{i,n}}} |}^2}{p_i}$. It is obvious that functions $E_k^{g - g,tra}\left( {{p_k}} \right)$ and $E_k^{g - s,tra}\left( {{p_k}} \right)$ are monotonically increasing for variable $p_k$, whose minimum values can be obtained when the variable is minimized. Thus, according to the delay requirements of GUEs in constraint (\ref{OP0st1}), the optimal solution for the transmit power of GUEs can be computed as
\begin{equation}\label{optimal p}
{p_k} = \left\{\begin{array}{l}
\min \left(\frac{{I^g_{k,m} + \delta _1^2} }{{{{| {w_{k,m}^H{h_{k,m}}} |}^2}}}( {{2^{\frac{{{d_k}}}{{{B_1}Z_{k,m}^{\alpha}}}}} - 1} ), P_k^{\max}\right),\text{if}\ {\alpha _{k,m}} = 1,\\
\min \left(\frac{{I^s_{k,n} + \delta _2^2} }{{{{| {v_{k,n}^H{g_{k,n}}} |}^2}}}( {{2^{\frac{{{d_k}}}{{{B_2}Z_{k,n}^{\beta}}}}} - 1} ), P_k^{\max}\right),\text{if}\ {\beta _{k,n}} = 1,
\end{array} \right.
\end{equation}
where $Z_{k,m}^{\alpha}={Z_k^g - {{{d_k}{c_k}}}/{{f_{k,m}^{gro}}}}$ and $Z_{k,n}^{\beta}={Z_k^g - {{{d_k}{c_k}}}/{{f_{k,n}^{sat-g}}}-{{{\varphi _{k,n}}}}/{\mu }}$. In addition, the transmission energy consumption from the $l$-th SUE to the LEO satellite is a nonconvex function for the transmit power of SUEs ${q_l}$, namely
\begin{equation}\label{E(q_l)}
E_l^{s - s,tra}\left( {{q_l}} \right) = {q_l}\frac{{d_l^{space}}}{{{B_3}{{\log }_2}\left( {1 + {q_l}{I_l}} \right)}},
\end{equation}
where ${I_l} = {{\sum\limits_{n = 1}^N {{\gamma _{l,n}}\eta _l^t\eta _n^r{{\left( {\frac{\lambda }{{4\pi {\phi _{l,n}}}}} \right)}^2}G_l^tG_n^rL_{l}^tL_{n}^r} }} / {{\delta _3^2}}$. In order to resolve the nonconvexity of $E_l^{s - s,tra}\left( {{q_l}} \right)$ in (\ref{E(q_l)}), we introduce auxiliary variable ${\tilde q_l} = \frac{1}{{{{\log }_2}\left( {1 + {q_l}{I_l}} \right)}} \ge 0$, and then $E_l^{s - s,tra}$ can be rewritten as a convex function of $\tilde{q}_l$, i.e.,
\begin{equation}\label{E(q_l)_convex}
E_l^{s - s,tra}\left( {{{\tilde q}_l}} \right) = \frac{{d_l^{space}}}{{{B_3}{I_l}}}{\tilde q_l}\left( {{2^{\frac{1}{{{{\tilde q}_l}}}}} - 1} \right).
\end{equation}
At the same time, the transmission delay $T_l^{s - s,tra}$ from the $l$-th SUE to the LEO satellite is replaced by
\begin{equation}\label{T(q_l)}
T_l^{s - s,tra}\left( {{{\tilde q}_l}} \right) = \frac{{{{{\tilde q}_l}d_l^{space}}}}{{{{B_3}}}}.
\end{equation}
For a given transmit power $p_k$ of GUEs based on (\ref{optimal p}), the original problem (\ref{OP3}) is equivalently transformed as
\begin{subequations}\label{OP3-1}
\begin{align}
\mathop {\min }\limits_{\tilde{\bf{q}},\bf{f}} & \sum\limits_{k = 1}^K {\rho _k^g( {E_k^{g - g} + E_k^{g - s}} )}  + \sum\limits_{l = 1}^L {\rho _l^s (E_l^{s - s,tra}( {{{\tilde q}_l}} )+ E_l^{s - s,com} ) }\label{OP3-1obj}\\
\textrm{s.t.}\
&(\ref{OP0st1}),(\ref{OP0st3}),(\ref{OP0st4}),\notag\\
&T_l^{s - s,tra}\left( {{{\tilde q}_l}} \right)+T_l^{s - s,com}+T_l^{s - s,pro} \le Z_l^s,\\
&{{\tilde q}_l} \ge \frac{1}{{{{\log }_2}\left( {1 + Q_l^{\max }{I_l}} \right)}},\label{OP3-1st1}
\end{align}
\end{subequations}
where $\tilde{\bf{q}}=\{\tilde{q}_l,\forall l \in L\}$. Since the objective function and all constraints of problem (\ref{OP3-1}) are convex, it is likely to obtain its optimal solution by off-the-shelf methods. After obtaining the solution, the transmit power of SUEs can be computed as
\begin{equation}\label{solution q}
{q_l} =\left( {{2^{\frac{1}{{\tilde q_l}}}} - 1} \right)/{I_l}.
\end{equation}
In conclusion, by iteratively optimizing offloading selection subproblem, beamforming design subproblem and resource allocation subproblem, a feasible solution can be achieved when its objective value converges. The proposed AO-based algorithm for satellite-terrestrial computing is summarized as Algorithm 1.

\emph{Remark 1:} Note that the integrated satellite-terrestrial computing framework proposed in this paper is suitable for dynamic satellite-terrestrial networks. In particular, the satellite-terrestrial network is dynamic over time slots, but is relatively static in a time slot. For a dynamic satellite-terrestrial computing networks, once the topology and channel information is obtained in a time slot, the proposed Algorithm 1 can be employed to carry out computing offloading.
As a result, different optimization results can be obtained for different time slots based on the dynamic characteristics of the network and channel to achieve highly reliable communication and low-latency computing at any moment. With strong payload processing capabilities of the new-generation LEO satellites, the computing task is able to be transmitted within a
single time slot. In addition, we plan to analyze multi-hop computing offloading in the subsequent research to adjust the routing plan in time according to the dynamic changes of the network topology and to select the best transmission paths to achieve the cooperative resources utilization and efficient computing tasks processing.

\begin{algorithm} %Algorithm 1
\caption{: Energy-Efficient Design of Satellite-terrestrial Computing}
\label{alg1}
\hspace*{0.02in} {\bf Input:} %算法的结果输入
 $K,L,M,N,N_t^g,N_t^s,Z_k^g,Z_l^s,d_k,c_k,d_l^{space},c_l^{space}\rho _k^g,$\\
  ${\ \ \ \ \ \ \ \ \ \ }\rho _l^s,F_m^{gro},F_n^{sat},P_k^{\max},Q_l^{\max} $.\\
\hspace*{0.02in} {\bf Output:} %算法的结果输出
$\bm{\alpha},\bm{\beta},\bm{\gamma},\bm{w},\bm{v},\bf{p},\bf{q},\bf{f}$.
\begin{algorithmic}[1]
\STATE{\textbf{Initialize} iteration index $t=0$, $p_k^{(t)}=P_k^{\max}/2$, $q_l^{(t)}=Q_l^{\max}/2$, $f_{k,m}^{gro(t)}=F_m^{gro}/K$, $f_{k,n}^{sat-g(t)}=F_n^{sat}/(K+L)$, $f_{l,n}^{sat-s(t)}=F_n^{sat}/(K+L)$.}
\REPEAT
\STATE{Obtain $\bm{\alpha}^{(t+1)}$, $\bm{\beta}^{(t+1)}$ and $\bm{\gamma}^{(t+1)}$ based on the relaxation mapping method according to the problem (\ref{OP1-1}) with fixed $\bm{w}^{(t)}$, $\bm{v}^{(t)}$, ${\bf{p}}^{(t)}$, ${\bf{q}}^{(t)}$ and ${\bf{f}}^{(t)}$;}
\STATE{Obtain $\bm{w}^{(t+1)}$ and $\bm{v}^{(t+1)}$ by solving problem (\ref{OP2-2}) with fixed $\bm{\alpha}^{(t+1)}$, $\bm{\beta}^{(t+1)}$, $\bm{\gamma}^{(t+1)}$, ${\bf{p}}^{(t)}$, ${\bf{q}}^{(t)}$ and ${\bf{f}}^{(t)}$;}
\STATE{Compute ${\bf{p}}^{(t+1)}$ by equation (\ref{optimal p});}
\STATE{Obtain ${\bf{q}}^{(t+1)}$ and ${\bf{f}}^{(t+1)}$ by solving problem (\ref{OP3-1}) with fixed $\bm{\alpha}^{(t+1)}$, $\bm{\beta}^{(t+1)}$, $\bm{\gamma}^{(t+1)}$, $\bm{w}^{(t+1)}$, $\bm{v}^{(t+1)}$ and ${\bf{p}}^{(t+1)}$;}
\STATE{Update $t=t+1$;}
\UNTIL Convergence
\end{algorithmic}
\end{algorithm}

\subsection{Algorithm Analysis}
In this part, we analyze the convergence and complexity of the proposed algorithm to verify its feasibility for satellite-terrestrial computing in 6G wireless networks.

\emph{Convergence Analysis:}
For the sake of description, the objective value of problem (\ref{OP0}) at the $t$-th iteration is defined as
$\Xi\left(\bm{\alpha}^{(t)}, \bm{\beta}^{(t)}, \bm{\gamma}^{(t)}, \bm{w}^{(t)}, \bm{v}^{(t)}, {\bf{p}}^{(t)}, {\bf{q}}^{(t)}, {\bf{f}}^{(t)}\right)$. According to Algorithm 1, the solution to the original problem (\ref{OP0}) is achieved by iteratively implementing steps 3-6. Particularly, in step 3 of Algorithm 1, since the optimal solutions of offloading selection subproblem $\bm{\alpha}^{(t+1)}$, $\bm{\beta}^{(t+1)}$ and $\bm{\gamma}^{(t+1)}$ are obtained based on the relaxation mapping method with given other variables, we have
\begin{align}\label{Convergence 1}
&\Xi\left(\bm{\alpha}^{(t)},\bm{\beta}^{(t)},\bm{\gamma}^{(t)},\bm{w}^{(t)},\bm{v}^{(t)}, {\bf{p}}^{(t)},{\bf{q}}^{(t)},{\bf{f}}^{(t)}\right)\notag\\
&\geq\Xi\left(\bm{\alpha}^{(t+1)}, \bm{\beta}^{(t+1)},\bm{\gamma}^{(t+1)},\bm{w}^{(t)},\bm{v}^{(t)},{\bf{p}}^{(t)}, {\bf{q}}^{(t)},{\bf{f}}^{(t)}\right).
\end{align}
Next, in step 4 of Algorithm 1, due to the convexity of beamforming design subproblem (\ref{OP2-2}), we obtain
\begin{align}\label{Convergence 2}
&\Xi\left(\bm{\alpha}^{(t+1)},\bm{\beta}^{(t+1)},\bm{\gamma}^{(t+1)},\bm{w}^{(t)},\bm{v}^{(t)}, {\bf{p}}^{(t)},{\bf{q}}^{(t)},{\bf{f}}^{(t)}\right)\notag\\
&\geq \Xi\left(\bm{\alpha}^{(t+1)}, \bm{\beta}^{(t+1)},\bm{\gamma}^{(t+1)},\bm{w}^{(t+1)},\bm{v}^{(t+1)},{\bf{p}}^{(t)}, {\bf{q}}^{(t)},{\bf{f}}^{(t)}\right).
\end{align}
Then, because of the monotonicity of the objective function for the transmit power of GUEs, it is known that
\begin{align}\label{Convergence 3}
&\Xi\left(\bm{\alpha}^{(t+1)},\bm{\beta}^{(t+1)},\bm{\gamma}^{(t+1)},\bm{w}^{(t+1)},\bm{v}^{(t+1)}, {\bf{p}}^{(t)},{\bf{q}}^{(t)},{\bf{f}}^{(t)}\right)\notag\\
&\geq \Xi\left(\bm{\alpha}^{(t+1)}, \bm{\beta}^{(t+1)},\bm{\gamma}^{(t+1)},\bm{w}^{(t+1)},\bm{v}^{(t+1)},{\bf{p}}^{(t+1)}, {\bf{q}}^{(t)},{\bf{f}}^{(t)}\right).
\end{align}
Similarly, since problem (\ref{OP3-1}) is jointly convex with respect to the transmit power of SUEs and computing power, we have
\begin{align}\label{Convergence 4}
&\Xi\left(\bm{\alpha}^{(t+1)},\bm{\beta}^{(t+1)},\bm{\gamma}^{(t+1)},\bm{w}^{(t+1)},\bm{v}^{(t+1)}, {\bf{p}}^{(t+1)},{\bf{q}}^{(t)},{\bf{f}}^{(t)}\right)\notag\\
&\geq \Xi\Big(\bm{\alpha}^{(t+1)}, \bm{\beta}^{(t+1)},\bm{\gamma}^{(t+1)},\bm{w}^{(t+1)},\bm{v}^{(t+1)},\notag\\
&\ \ \ \ \ \ \ \ \ \ \ \ \ \ \ \ \ \ \ \ \ \ \ \ \ \ \ \ \ \ \ \ \ \ \ \ {\bf{p}}^{(t+1)}, {\bf{q}}^{(t+1)},{\bf{f}}^{(t+1)}\Big).
\end{align}
Finally, based on (\ref{Convergence 1}), (\ref{Convergence 2}), (\ref{Convergence 3}) and (\ref{Convergence 4}), we can conclude
\begin{align}\label{Convergence}
&\Xi\left(\bm{\alpha}^{(t)},\bm{\beta}^{(t)},\bm{\gamma}^{(t)},\bm{w}^{(t)},\bm{v}^{(t)}, {\bf{p}}^{(t)},{\bf{q}}^{(t)},{\bf{f}}^{(t)}\right)\notag\\
&\geq \Xi\Big(\bm{\alpha}^{(t+1)}, \bm{\beta}^{(t+1)},\bm{\gamma}^{(t+1)},\bm{w}^{(t+1)},\bm{v}^{(t+1)},\notag\\
&\ \ \ \ \ \ \ \ \ \ \ \ \ \ \ \ \ \ \ \ \ \ \ \ \ \ \ \ \ \ \ \ \ \ \ \ {\bf{p}}^{(t+1)}, {\bf{q}}^{(t+1)},{\bf{f}}^{(t+1)}\Big),
\end{align}
which indicates that the weighted total energy consumption is non-increasing during the iterations of Algorithm 1. In addition, there is a lower bound on the weighted total energy consumption of the satellite-terrestrial computing system, owing to the delay requirements of computing tasks. As a result, the convergence of Algorithm 1 is guaranteed according to the monotone bounded convergence theorem \cite{Convergence analyse}. Furthermore, we confirm the convergence of Algorithm 1 for different numbers of GUEs and SUEs by simulation in Fig. \ref{temp_E_K}.

\emph{Complexity Analysis:}
It is worth pointing out that the computational complexity of Algorithm 1 is mainly attributed to step 3, step 4 and step 6. For linear programming or convex optimization problem which only contains linear matrix inequalities (LMI) and second-order cone (SOC) constraints can be effectively solved by the interior point method \cite{IPM}. Thus, it is possible to measure the worst-case complexity of the proposed algorithm by the interior point method \cite{complexity analysis}. Specifically, for the relaxation optimization problem (20) in step 3, it has $2K+2L+M+N$ LMI constraints of dimension 1. Thus, for a given precision $\zeta_1 > 0$, the worst-case complexity of obtaining the optimal solution for problem (20) is $\sqrt{2K+2L+M+N}\cdot\varpi_1\cdot\ln(1/\zeta_1)$, where $\varpi_1=z_1[2K+2L+M+N+z_1(2K+2L+M+N)+z_1^2]$
with decision variable $z_1=\mathcal{O}(KM+KN+LN)$.
Similarly, for the beamforming design subproblem (31) in step 4, there are $K(3M+3N+1)$ LMI constraints of dimension 1, $KM$ LMI constraints of dimension $N_t^g$, $KN$ LMI constraints of dimension $N_t^s$, $KM$ SOC constraints of dimension $N_t^g+1$ and $KN$ SOC constraints of dimension $N_t^s+1$. In this context, for a given precision $\zeta_2 > 0$, the worst-case complexity of obtaining the optimal solution for problem (31) is $\sqrt{K(M(N_t^g+5)+N(N_t^s+5)+1)}\cdot\varpi_2\cdot\ln(1/\zeta_2)$, where $\varpi_2=z_2[KM((N_t^g)^3+(N_t^g+1)^2+3)+KN((N_t^s)^3+(N_t^g+1)^2+3)+K+Kz_2(M((N_t^g)^2+3)+N((N_t^s)^2+3)+1)+z_2^2]$
with decision variable $z_2=\mathcal{O}( {KM{{(N_t^g)}^2} + KN{{(N_t^s)}^2}} )$.
Finally, for step 6, there are $K+2L+M+N$ LMI constraints of dimension 1. Thus, for a given precision $\zeta_3 > 0$, the worst-case complexity of solving the problem (39) is $\sqrt{K+2L+M+N}\cdot\varpi_3\cdot\ln(1/\zeta_3)$, where $\varpi_3=z_3[K+2L+M+N+z_3(K+2L+M+N)+z_3^2]$
with decision variable $z_3=\mathcal{O}(KM+KN+LN)$.
To make it more intuitive, we show the realistic runtime of the proposed algorithm in an Intel i5-10400F CPU by MATLAB simulation for different parameters in Table I. It is worth noting that dedicated processors are used for parallel computing processing in real-world applications to achieve millisecond response.
\begin{table*}[ht]
\small
\centering
\caption{The Realistic Runtime (s) of Algorithm 1}\label{runtime}
\begin{tabular}{|c|c|c|c|c|c|c|c|c|}
\hline
$N_t^g=N_t^s$ & 16 & 24 & 32 & 40 & 48 & 56 & 64 \\\hline
$K=10, M=N=2$ & 1.4088 & 1.5539 & 1.7751 & 2.0879 & 2.5873 & 3.1910 & 4.3601 \\\hline
$K=20, M=N=2$ & 2.7269 & 3.2337 & 3.9893 & 5.4930 & 8.1524 & 10.5377 & 12.8983 \\\hline
$K=10, M=N=4$ & 2.0136 & 2.2904 & 2.7164 & 3.4456 & 4.3796 & 6.2297 & 7.5781 \\\hline
$K=20, M=N=4$ & 4.1305 & 5.1796 & 7.2161 & 11.1879 & 15.1425 & 18.7489 & 23.8465 \\\hline
\end{tabular}
\end{table*}

\section{Simulation Results}
\begin{table*}[ht]
\small
\centering
\caption{Simulation Parameters }\label{Simulation}
\begin{tabular}{|c|c|}
\hline
Parameters & Values \\ \hline
Number of GUEs and SUEs & $K=10, L=10$ \\\hline
Number of BSs and its antennas & $M=2, N_t^g=16$ \\\hline
Number of LEO satellites and its antennas & $N=3, N_t^s=16$ \\\hline
Bandwidth & $B_1=B_2=20$ MHz, $B_3=100$ MHz \\\hline
Maximum transmit power budget of GUE and SUE  & $P_0^{\max}=30$ dBm , $Q_0^{\max}=30$ dBm \\\hline
Computing task of GUE & $d_k\in[200\sim400]$ KB, $c_k\in[100\sim150]$ cycles/bit  \\\hline
Computing task of SUE & $d_k^{space}\in[200\sim400]$ KB, $c_k^{space}\in[100\sim150]$ cycles/bit  \\\hline
Delay requirement of GUE and SUE & $Z_0^g=100$ ms, $Z_0^s=100$ ms \\\hline
Maximum computing power of BS and LEO satellite & $F_0^{gro}=30$ GHz, $F_0^{sat}=10$ GHz \\\hline
Energy weights for GUE and SUE & $\rho_k^g=1$, $\rho_l^s=1$ \\\hline
Noise power & $\sigma_1^2=\sigma_2^2=\sigma_3^2=-110$ dBm  \\\hline
Boltzmann constant & $\kappa=1.38 \times 10^{-23}$ J/m  \\\hline
Energy coefficient & $\tau _m^{gro}=\tau _n^{sat}= 5\times10^{-27}$ \cite{tau} \\\hline
Carrier frequency & $f=6$ GHz  \\\hline
Distance between SUE and LEO satellite & $\phi_{l,n}\in[500-1500]$ km \\\hline
Transmit antenna gain per noise temperature & $G_{k,n}/T=34$ dB/K \\\hline
Rain fading mean and variance & $\mu_r=-2.6$ dB, $\sigma_r^2=1.63$ dB \\\hline
3-dB angle & $\varepsilon_n^{3dB}=0.4^{\circ}$ \\\hline
Maximum satellite antenna gain & $b_{n,\max}=14$ dBi \\\hline
Optical efficiency of the transmitter and receiver & $\eta_l^t=0.9$, $\eta_n^r=0.9$ \\\hline
Wavelength & $\lambda=1550$ nm \\\hline
Aperture diameter of transmitter and receiver & $D_l^t=20$ cm, $D_n^r=20$ cm \\\hline
Pointing error angle of transmitter and receiver & $e_{l}^t=0.8\ \mu$rad, $e_{n}^r=0.8\ \mu$rad \\\hline
\end{tabular}
\end{table*}

In this section, we present simulation results to validate the effectiveness of the proposed algorithm in practical satellite-terrestrial computing systems. Particularly, we consider that there is a Walker Delta constellation with orbital altitude of 550 km, inclination of 53 degrees, and constellation parameters of 1584/72/1, which represents the number of planes as 72, with 22 satellites in each plane and the phase factor of 1 \cite{Walker Delta1}. Then, we select a set of neighboring satellites in the above satellite constellation for simulation based on the desired number of LEO satellites. In this context, it can be obtained that the communication distance $\varphi_{k,n}$ between the GUE and the LEO satellite is in the range from about 550 km to 2700 km \cite{Walker Delta2}. For convenience, it is assumed that all GUEs/SUEs have the same maximum transmission power budgets and delay requirements, and all MEC servers at the BSs/LEO satellites have the same maximum computing power, i.e., $P_k^{\max}=P_0^{\max}$, $Q_l^{\max}=Q_0^{\max}$, $Z_k^g=Z_0^g$, $Z_l^s=Z_0^s$, $F_m^{gro}=F_0^{gro}$ and $F_n^{sat}=F_0^{sat}$, $\forall k,l,m,n$. Unless otherwise stated, the default simulation parameters are listed in Table II.

\begin{figure}[h]
 \centering
\includegraphics [width=0.45\textwidth] {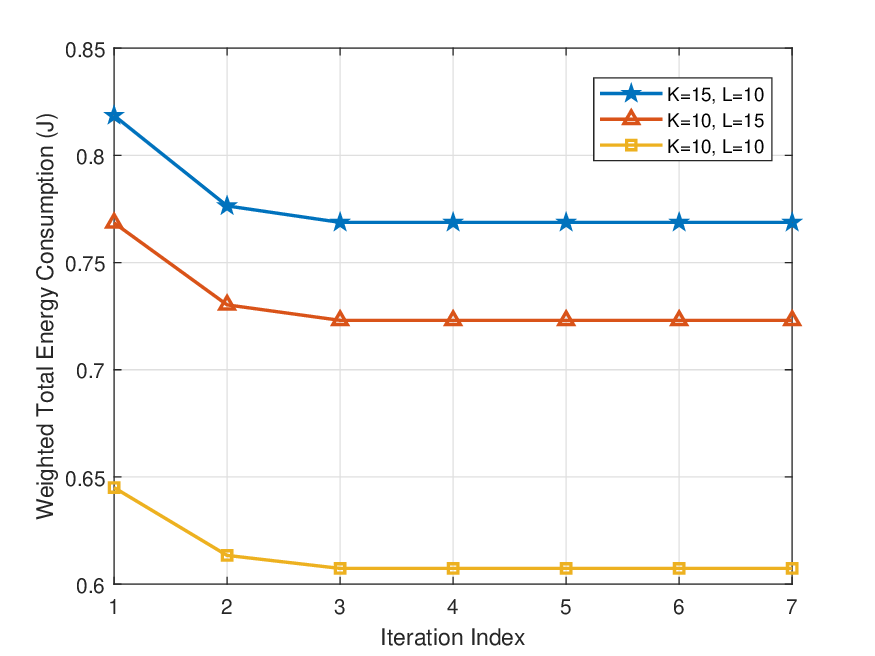}
\caption {Convergence behavior of Algorithm 1.}
\label{temp_E_K}
\end{figure}
First of all, we provide the convergence of the Algorithm 1 with different numbers of GUEs and SUEs. From Fig. \ref{temp_E_K}, it is evident that the value of the weighted total energy consumption decreases monotonically over the iterations and converges to a stable point within 5 iterations. Thus, the computational complexity of the proposed algorithm is affordable in practical applications for satellite-terrestrial computing systems.

\begin{figure}[h]
 \centering
\includegraphics [width=0.45\textwidth] {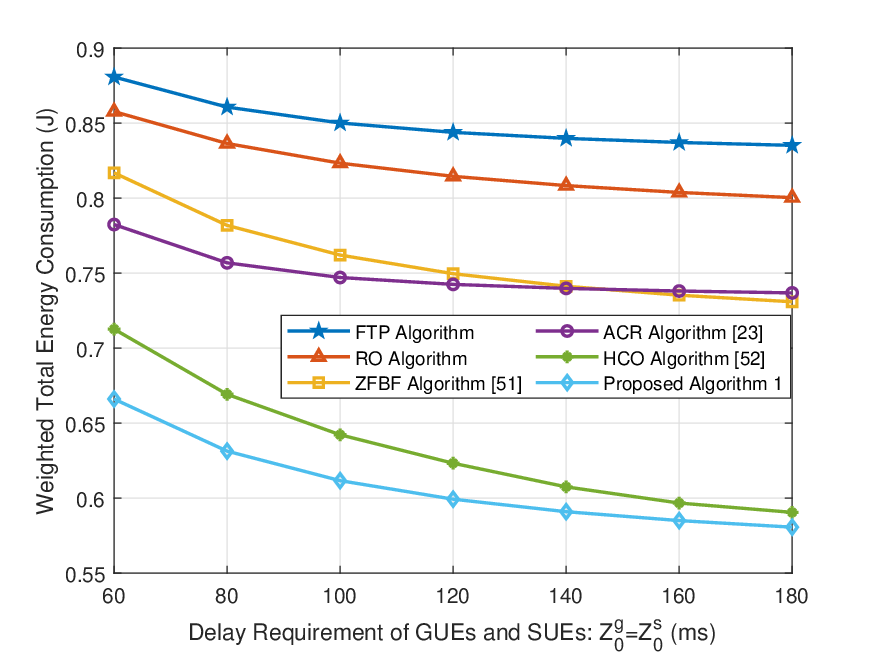}
\caption {Performance comparison of different algorithms.}
\label{compare}
\end{figure}
Then, we present the superior performance of Algorithm 1 compared to five baseline algorithms , i.e., Fixed Transmit Power (FTP) Algorithm with $p_k=P_K^{\max}/2$ and $q_l=Q_l^{\max}/2$, Zero-Forcing Beamforming (ZFBF) Algorithm with zero-forcing receivers on BSs and LEO satellites \cite{ZFBF}, Random Offloading (RO) Algorithm by randomly selecting any BS or LEO satellite, Average Computing Resources (ACR) Algorithm with $f_{k,m}^{gro}=F_m^{gro}/K$ and $f_{k,n}^{sat-g}=f_{l,n}^{sat-s}=F_n^{sat}/(K+L)$ \cite{related work 3} and Heuristic Computing Offloading (HCO) Algorithm based on the constrained particle swarm optimization proposed in the related work \cite{HCO}.
In Fig. \ref{compare}, it is seen that the proposed Algorithm 1 always consumes the minimum weighted total energy compared to other algorithms in the whole delay requirement region. This can be attributed to the fact that the proposed Algorithm 1 is adaptively optimized according to the characteristics of the integrated satellite-terrestrial 6G wireless network compared to other baselines, and heuristic algorithms often fail to reach the optimal value when solving large-scale 0-1 programming problems, which also validates the effectiveness of the proposed Algorithm 1.
In addition, the more stringent delay requirement of GUEs and SUEs, the higher the energy consumption. This is because higher transmit power is used for data transmission and more computing resources are used to complete the computing tasks to meet stringent delay requirements. Thus, it makes sense to balance the performance between time cost and energy cost based on realistic scenarios.

\begin{figure}[h]
 \centering
\includegraphics [width=0.45\textwidth] {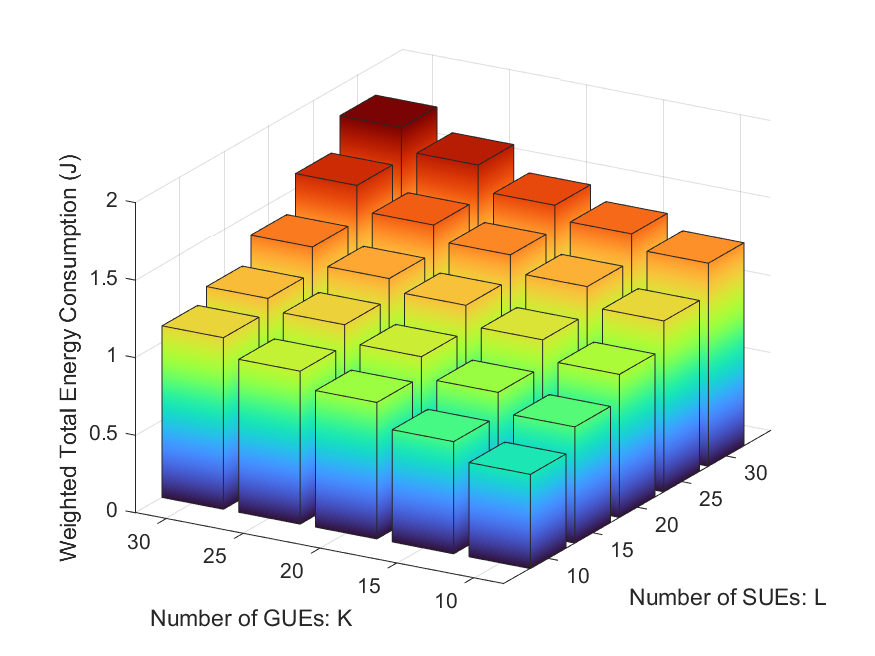}
\caption {Weighted total energy consumption versus different numbers of GUEs and SUEs.}
\label{K_L_E}
\end{figure}
Next, Fig. \ref{K_L_E} reveals the impacts of the number of GUEs $K$ and the number of SUEs $L$ on the weighted total energy consumption for satellite-terrestrial computing in 6G wireless networks. Apparently, as the number of GUEs and SUEs increases, the weighted total energy consumption increases accordingly. On the one hand, the more data need to be processed, the more energy caused by data transmission and computing is consumed. On the other hand, increasing the number of GUEs will bring more co-channel interference, which degrades the data transmission efficiency. Therefore, the number of GUEs and SUEs supported by the integrated satellite-terrestrial 6G wireless network should match its affordable energy consumption level.

\begin{figure}[h]
 \centering
\includegraphics [width=0.45\textwidth] {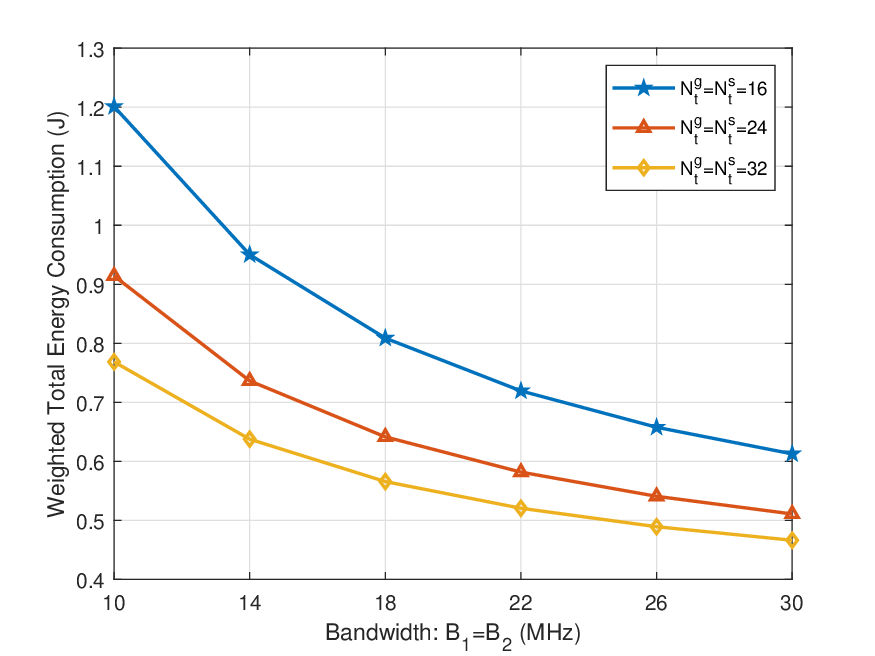}
\caption {Weighted total energy consumption versus bandwidth for different numbers of antennas at BSs and LEO satellites.}
\label{B_E_Nt}
\end{figure}
Furthermore, we investigate the effects of the bandwidth of the inter-terrestrial and satellite-terrestrial channels, as well as the number of antennas at BSs and LEO satellites on the weighted total energy consumption. In Fig. \ref{B_E_Nt}, it is seen that the weighted total energy consumption decreases as the channel bandwidth and the number of antennas increase, because both of them have a significant impact on the data transmission rate for computing tasks. Notice that the performance gain from $N_t^g=N_t^s=24$ to $N_t^g=N_t^s=32$ is less than that from $N_t^g=N_t^s=16$ to $N_t^g=N_t^s=24$, indicating that the performance gain of the system by increasing the number of antennas is limited. Thus, a suitable number of antennas should be deployed at the BSs and LEO satellites to balance performance and cost.

\begin{figure}[h]
 \centering
\includegraphics [width=0.45\textwidth] {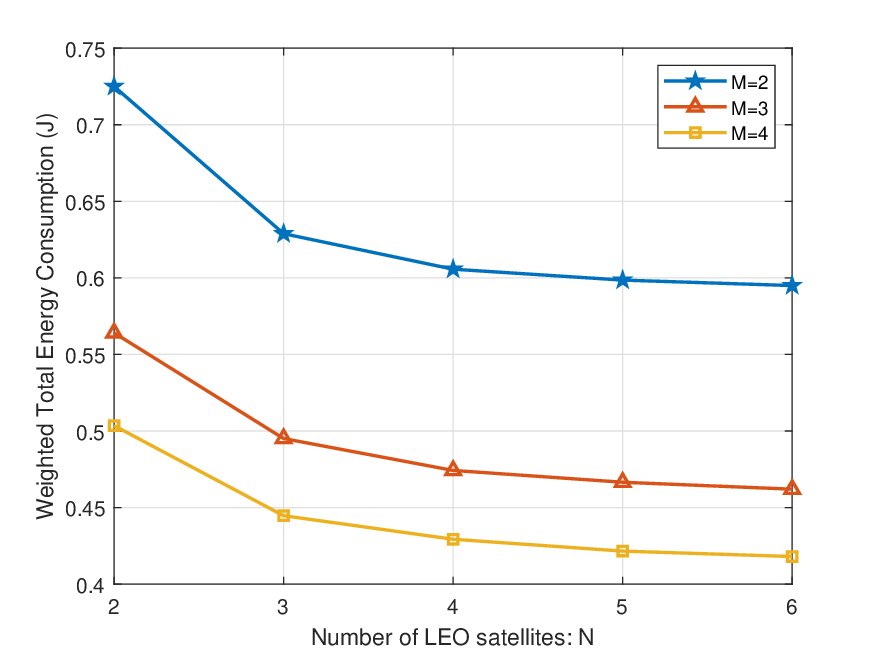}
\caption {Weighted total energy consumption versus number of LEO satellites for different numbers of BSs.}
\label{N_E_M}
\end{figure}
Fig. \ref{N_E_M} examines the influence of the number of BSs $M$ and the number of LEO satellites $N$ for satellite-terrestrial computing in 6G wireless networks. As is expected that increasing the number of BSs and LEO satellites leads to lower energy consumption, since it expands the total computing power of the system and also gives more offloading options for computing tasks at the same time.
In fact, the proposed Algorithm 1 theoretically supports an arbitrary number of BSs and LEO satellites. However, in practical systems, it is essential to consider a combination of deployment costs, operational costs, and actual user requirements. The appropriate number of BSs and LEO satellites should be determined based on the specific circumstances to ensure the feasibility and cost-effectiveness of the system and to achieve the best possible performance in practical applications. Meanwhile, with the development of lightweight satellites, construction and launch costs have been greatly reduced, making it possible to deploy a large number of LEO satellites in real-world applications, e.g., starlink program with 42000 LEO satellites.

\begin{figure}[h]
 \centering
\includegraphics [width=0.45\textwidth] {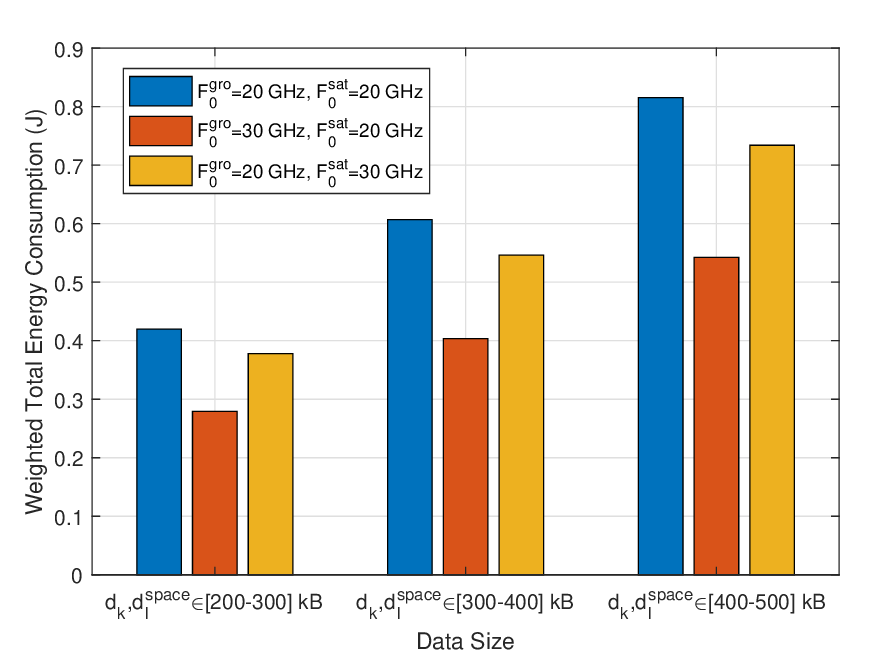}
\caption {Weighted total energy consumption versus data sizes of computing tasks for different maximum computing power.}
\label{d_E_F}
\end{figure}
Finally, we show the impacts of the data size of computing tasks and the maximum computing power provided by the MEC servers on the performance of the weighted total energy consumption. It is seen that a larger amount of data requires more energy consumption since it requires longer time both in data transmission and processing. Moreover, the increase of the maximum computing capacity of the MEC servers at the BSs saves more energy consumption than that at the LEO satellites. This is because the computing tasks of GUEs are preferentially offloaded to the BSs when their MEC severs have sufficient computing resources, in order to avoid the long-distance propagation loss caused by the satellite-terrestrial communication.

\section{Conclusion}
This paper presented a comprehensive satellite-terrestrial computing architecture in 6G wireless networks. To enhance the overall performance, an energy-efficient design was proposed according to the characteristics of integrated satellite-terrestrial networks. In particular, the design was formulated as a complicated MINLP problem with weighted total energy consumption minimization while ensuring the delay requirements of GUEs and SUEs. To obtain the feasible solution, we decompose the original NP-hard problem into three subproblems, i.e., offloading selection, beamforming design and resource allocation, and then solve them iteratively in turn. Theoretical analysis confirmed the fast convergence behavior and low computational complexity of the proposed algorithm. Moreover, simulation results revealed some useful insights of parameter selection of the proposed algorithm.

\begin{appendices}
\section{Proof of the Rank-one Constraints}
Herein, we give the proof for dropping the rank-one constraints in problem (\ref{OP2-2}). Based on the offloading selection, problem (\ref{OP2-2}) has two cases that the computing tasks are offloaded to the BS or to the LEO satellite. In particular, when $\alpha_{k,m}=1$, the Lagrangian function for problem (\ref{OP2-2}) with respect to $\mathbf{W}_{k,m}$ can be obtained as
\begin{align*}
&\mathcal{L}(\mathbf{W}_{k,m})=\rho_k^g{p_k}A_{k,m}^g+\xi_1\left(\text{tr}(\mathbf{W}_{k,m})-1\right)\\
&\ \ \ \ \ \ +\xi_2\left(A_{k,m}^g+\frac{{d_k}{c_k}}{f_{k,n}^{sat - g}}-Z_k^g\right)+\xi_3\left(\frac{{{d_k}}}{{\tilde R_{k,m}^{g - g}}}-A_{k,m}^g\right)\\
&\ \ \ \ \ \ +\xi_4\left(\tilde R_{k,m}^{g - g}-{B_1}{\log _2}\left( {1 + \tilde \Gamma _{k,m}^{g - g}} \right)\right)\\
&\ \ \ \ \ \ +\xi_5\Bigg(\sum\limits_{i = k + 1}^K {\text{tr}({\mathbf{h}_{i,m}}\mathbf{h}_{i,m}^H{\mathbf{W}_{k,m}}){p_i} + \delta _1^2}\\
&\ \ \ \ \ \ \ \ \ \ \ \ -\frac{{\text{tr}({\mathbf{h}_{k,m}}\mathbf{h}_{k,m}^H{\mathbf{W}_{k,m}}){p_k}}}{{\tilde \Gamma _{k,m}^{g - g}}}\Bigg)-\mathbf{\Omega}\mathbf{W}_{k,m},
\end{align*}
where $\xi_1, \ldots, \xi_5$ and $\mathbf{\Omega}$ are the Lagrange multipliers. By utilizing the Karush-Kuhn-Tucher (KKT) conditions, we have
\begin{subequations}
\begin{align}
&\sum\limits_{i = k + 1}^K {\text{tr}({\mathbf{h}_{i,m}}\mathbf{h}_{i,m}^H{\mathbf{W}_{k,m}^*}){p_i} + \delta _1^2}\leq
\frac{{\text{tr}({\mathbf{h}_{k,m}}\mathbf{h}_{k,m}^H{\mathbf{W}_{k,m}^*}){p_k}}}{{\tilde \Gamma _{k,m}^{g - g}}}\label{KKT1}\\
&\mathbf{\Omega}\mathbf{W}_{k,m}^*=\mathbf{0}\label{KKT2}\\
&\nabla_{\mathbf{W}_{k,m}^*}\mathcal{L}=\xi_1^*\mathbf{I}_{N_t^g} \notag\\
&\ \ \ \ \ +\xi_5^*\left(\sum\limits_{i = k + 1}^K {{\mathbf{h}_{i,m}}\mathbf{h}_{i,m}^H{p_i}}-
\frac{{{\mathbf{h}_{k,m}}\mathbf{h}_{k,m}^H{p_k}}}{{\tilde \Gamma _{k,m}^{g - g}}}\right)-\mathbf{\Omega}^*=\mathbf{0}\label{KKT3}\\
&\mathbf{\Omega}^*\succeq\mathbf{0}, \mathbf{W}_{k,m}^*\succeq\mathbf{0}, \xi_1^*\geq0, \xi_5^*\geq0.\label{KKT4}
\end{align}
\end{subequations}
Since $\delta _1^2>0$, it is inferred from (\ref{KKT1}) that $\mathbf{W}_{k,m}^*\neq\mathbf{0}$, i.e.,
\begin{equation}\label{AP1}
\mathrm{Rank}(\mathbf{W}_{k,m}^*)\geq1.
\end{equation}
Moreover, it can be found from (\ref{KKT2}) that
\begin{equation}\label{AP2}
\mathrm{Rank}(\mathbf{\Omega}^*)+\mathrm{Rank}(\mathbf{W}_{k,m}^*)\leq{N_t^g}.
\end{equation}
Based on (\ref{AP1}) and (\ref{AP2}), we have
\begin{equation}\label{AP3}
\mathrm{Rank}(\mathbf{\Omega}^*)\leq{N_t^g-1}.
\end{equation}
According to the properties of the rank of matrix, it is known from (\ref{KKT3}) that
\begin{equation}
\mathrm{Rank}(\mathbf{\Omega}^*)+\mathrm{Rank}(\mathbf{\Upsilon})\geq\mathrm{Rank}(\xi_1^*\mathbf{I}_{N_t^g}),
\end{equation}
where $\mathbf{\Upsilon}=\xi_5^*\left(
\frac{{{\mathbf{h}_{k,m}}\mathbf{h}_{k,m}^H{p_k}}}{{\tilde \Gamma _{k,m}^{g - g}}}-\sum\limits_{i = k + 1}^K {{\mathbf{h}_{i,m}}\mathbf{h}_{i,m}^H{p_i}}\right)$.
Obviously, $\mathbf{\Upsilon}\neq\mathbf{0}$, namely $\mathrm{Rank}(\mathbf{\Upsilon})\geq1$. Considering the fact that $\mathrm{Rank}(\xi_1^*\mathbf{I}_{N_t^g})=N_t^g$, we can infer that
\begin{equation}\label{AP4}
\mathrm{Rank}(\mathbf{\Omega}^*)\geq{N_t^g-1}.
\end{equation}
Combining (\ref{AP3}) and (\ref{AP4}), we have $\mathrm{Rank}(\mathbf{\Omega}^*)={N_t^g-1}$. Substituting it into (\ref{AP2}), we can get
\begin{equation}\label{AP5}
\mathrm{Rank}(\mathbf{W}_{k,m}^*)\leq1.
\end{equation}
Finally, it is proved from (\ref{AP1}) and (\ref{AP5}) that $\mathrm{Rank}(\mathbf{W}_{k,m}^*)=1$. Similarly, when $\beta_{k,n}=1$, it also can be verified that $\mathrm{Rank}(\mathbf{V}_{k,n}^*)=1$ always holds true. The proof is finished.
\end{appendices}


\begin{thebibliography}{1}

\bibitem{MEC}
K. Dolui and S. K. Datta, ``Comparison of edge computing implementations: Fog computing, cloudlet and mobile edge computing," in \emph{Proc. Glob. Internet Things Summit}, 2017, pp. 1-6.

\bibitem{statistics}
C. Ding, J. -B. Wang, H. Zhang, M. Lin, and G. Y. Li, ``Joint optimization of transmission and computation resources for satellite and high altitude platform assisted edge computing," \emph{IEEE Trans. Wireless Commun.}, vol. 21, no. 2, pp. 1362-1377, Feb. 2022.

\bibitem{6G satellite1}
X. Zhu and C. Jiang, ``Integrated satellite-terrestrial networks toward 6G: Architectures, applications, and challenges," \emph{IEEE Internet Things J.}, vol. 9, no. 1, pp. 437-461, Jan. 2022.

\bibitem{6G satellite2}
X. Fang, W. Feng, T. Wei, Y. Chen, N. Ge, and C. -X. Wang, ``5G embraces satellites for 6G ubiquitous IoT: Basic models for integrated satellite terrestrial networks," \emph{IEEE Internet Things J.}, vol. 8, no. 18, pp. 14399-14417, Sep. 2021.

\bibitem{6G satellite3}
S. Chen, S. Sun, and S. Kang, ``System integration of terrestrial mobile communication and satellite communication -the trends, challenges and key technologies in B5G and 6G," \emph{China Commun.}, vol. 17, no. 12, pp. 156-171, Dec. 2020.

\bibitem{LEO1}
L. You, X. Qiang, C. G. Tsinos, F. Liu, W. Wang, X. Gao, and B. Ottersten, ``Beam squint-aware integrated sensing and communications for hybrid massive MIMO LEO satellite systems," \emph{IEEE J. Sel. Areas Commun.}, vol. 40, no. 10, pp. 2994-3009, Oct. 2022

\bibitem{LEO2}
Y. Zhang, Y. Wu, A. Liu, X. Xia, T. Pan, and X. Liu, ``Deep learning-based channel prediction for LEO satellite massive MIMO communication system," \emph{IEEE Wireless Commun. Lett.}, vol. 10, no. 8, pp. 1835-1839, Aug. 2021.

\bibitem{spacex}
J. Huang and J. Cao, ``Recent development of commercial satellite communications systems," in \emph{Proc. Artif. Intell. China}, 2020, pp. 531-536.

\bibitem{satellite-terrestrial network}
J. Zhang, X. Zhang, P. Wang, L. Liu, and Y. Wang, ``Double-edge intelligent integrated satellite terrestrial networks," \emph{China Commun.}, vol. 17, no. 9, pp. 128-146, Sep. 2020.

\bibitem{related work 0}
X. Zhu and C. Jiang, ``Delay optimization for cooperative multi-tier computing in integrated satellite-terrestrial networks," \emph{IEEE J. Sel. Areas Commun.}, vol. 41, no. 2, pp. 366-380, Feb. 2023.

\bibitem{related work 1}
Y. Song, X. Li, H. Ji, and H. Zhang, ``Joint computing, caching and communication resource allocation in the satellite-terrestrial integrated network with UE cooperation," in \emph{Proc. IEEE Int. Conf. Commun. China}, 2022, pp. 604-609.

\bibitem{add share spectral}
K. An, M. Lin, J. Ouyang, and W. -P. Zhu, ``Secure transmission in cognitive satellite terrestrial networks," \emph{IEEE J. Sel. Areas Commun.}, vol. 34, no. 11, pp. 3025-3037, Nov. 2016.

%\bibitem{OMA}
%F. Zou, R. Chai, and Q. Chen, ``Performance analysis of hybrid satellite-terrestrial relay networks," in \emph{Proc. IEEE Int. Symp. Person. Indoor Mobile Radio Commun.}, 2021, pp. 471-476.

\bibitem{OMA1}
A. Agarwal and P. Kumar, ``Analysis of variable bit rate SOFDM transmission scheme over multi-relay hybrid satellite-terrestrial system in the presence of CFO and phase noise," \emph{IEEE Trans. Veh. Technol.}, vol. 68, no. 5, pp. 4586-4601, May 2019.

\bibitem{OMA2}
J. Mashino and T. Sugiyama, ``Subcarrier suppressed transmission for OFDMA in satellite/terrestrial integrated mobile communication system," in \emph{Proc. IEEE Int. Conf. Commun.}, 2011, pp. 1-5.

\bibitem{related work 6}
Z. Lin, M. Lin, J. -B. Wang, T. de Cola, and J. Wang, ``Joint beamforming and power allocation for satellite-terrestrial integrated networks with non-orthogonal multiple access," \emph{IEEE J. Sel. Top. Signal Process.}, vol. 13, no. 3, pp. 657-670, Jun. 2019.

\bibitem{related work 2}
Y. Zhang, L. Yin, C. Jiang, and Y. Qian, ``Joint beamforming design and resource allocation for terrestrial-satellite cooperation system," \emph{IEEE Trans. Commun.}, vol. 68, no. 2, pp. 778-791, Feb. 2020.

\bibitem{add RSMA}
Z. Lin, M. Lin, T. de Cola, J. -B. Wang, W. -P. Zhu, and J. Cheng, ``Supporting IoT with rate-splitting multiple access in satellite and aerial-integrated networks," \emph{IEEE Internet Things J.}, vol. 8, no. 14, pp. 11123-11134, Jul. 2021.

\bibitem{add RIS}
Z. Lin, H. Niu, K. An, Y. Wang, G. Zheng, S. Chatzinotas, and Y. Hu, ``Refracting RIS aided hybrid satellite-terrestrial relay networks: Joint beamforming design and optimization," \emph{IEEE Trans. Aerosp. Electron. Syst.}, vol. 58, no. 4, pp.3717-3724, Aug. 2022.

\bibitem{add millimeter wave}
Z. Lin, M. Lin, B. Champagne, W. -P. Zhu, and N. Al-Dhahir, ``Secrecy-energy efficient hybrid beamforming for satellite-terrestrial integrated networks," \emph{IEEE Trans. Commun.}, vol. 69, no. 9, pp. 6345-6360, Sep. 2021.

\bibitem{TerrestrialComputing}
Q. Wang, X. Chen, and Q. Qi, ``Task-driven robust integration of communication and computation for edge-intelligent networks," \emph{IEEE Trans. Commun.}, vol. 71, no. 1, pp. 244-255.

\bibitem{offloading 1}
N. Eshraghi and B. Liang, ``Joint offloading decision and resource allocation with uncertain task computing requirement," in \emph{Proc. IEEE INFOCOM}, 2019, pp. 1414-1422.

\bibitem{offloading 2}
H. Guo, J. Zhang, J. Liu, and H. Zhang, ``Energy-aware computation offloading and transmit power allocation in ultradense IoT networks," \emph{IEEE Internet Things J.}, vol. 6, no. 3, pp. 4317-4329, Jun. 2019.

\bibitem{related work 3}
Y. Wang, J. Zhang, X. Zhang, P. Wang, and L. Liu, ``A computation offloading strategy in satellite terrestrial networks with double edge computing," in \emph{Proc. IEEE Int. Conf. Commun. Syst.}, 2018, pp. 450-455.

\bibitem{related work 4}
B. Wang, X. Li, D. Huang, and J. Xie, ``A profit maximization strategy of MEC resource provider in the satellite-terrestrial double edge computing system," in \emph{Proc. Int. Conf. Commun. Technol.}, 2021, pp. 906-912.

\bibitem{related work 5}
K. Wei, Q. Tang, J. Guo, M. Zeng, Z. Fei, and Q. Cui, ``Resource scheduling and offloading strategy based on LEO satellite edge computing," in \emph{Proc. IEEE Veh. Technol. Conf.}, 2021, pp. 1-6.

\bibitem{computing results}
Y. Mao, J. Zhang, and K. B. Letaief, ``Dynamic computation offloading for mobile-edge computing with energy harvesting devices," \emph{IEEE J. Sel. Areas Commun.}, vol. 34, no. 12, pp. 3590-3605, Dec. 2016.

\bibitem{time slots}
Q. Qi, X. Chen, and D. W. K. Ng, ``Robust beamforming for NOMA-based cellular massive IoT with SWIPT," \emph{IEEE Trans. Signal Process}., vol. 68, pp. 211-224, 2020.

\bibitem{Rayleigh fading}
Z. Song, Y. Hao, Y. Liu, and X. Sun, ``Energy-efficient multiaccess edge computing for terrestrial-satellite internet of things," \emph{IEEE Internet Things J.}, vol. 8, no. 18, pp. 14202-14218, Sep. 2021.

\bibitem{Pass loss}
3GPP, ``Coordinated multi-point operation for LTE physical layer aspects (Rel. 11)," Feb. 2011.

\bibitem{SIC}
Q. Gao, M. Jia, Q. Guo, X. Gu, and L. Hanzo, ``Jointly optimized beamforming and power allocation for full-duplex cell-free NOMA in space-ground integrated networks," \emph{IEEE Trans. Commun.}, vol. 71, no. 5, pp. 2816-2830, May 2023.

\bibitem{satellite-terrestrial channel}
Z. Gao, A. Liu, C. Han, and X. Liang, ``Max completion time optimization for internet of things in LEO satellite-terrestrial integrated networks," \emph{IEEE Internet Things J.}, vol. 8, no. 12, pp. 9981-9994, Jun. 2021.

\bibitem{satellite-terrestrial channel2}
J. Chu and X. Chen, ``Robust design for integrated satellite-terrestrial internet of things," \emph{IEEE Internet Things J.}, vol. 8, no. 11, pp. 9072-9083, Jun. 2021.

\bibitem{rain attenuation}
G. Zheng, S. Chatzinotas, and B. Ottersten, ``Generic optimization of linear precoding in multibeam satellite systems," \emph{IEEE Trans. Wireless Commun.}, vol. 11, no. 6, pp. 2308-2320, Jun. 2012.

\bibitem{beam gain}
M. A. Diaz, N. Courville, C. Mosquera, G. Liva, and G. E. Corazza, ``Non-Linear interference mitigation for broadband multimedia satellite systems," in \emph{Proc. Int. Workshop Satell. Space Commun.}, 2007, pp. 61-65.

\bibitem{Doppler}
L. You, K. -X. Li, J. Wang, X. Gao, X. -G. Xia, and B. Ottersten, ``Massive MIMO transmission for LEO satellite communications," \emph{IEEE J. Sel. Areas Commun.}, vol. 38, no. 8, pp. 1851-1865, Aug. 2020.

\bibitem{FSO 1}
P. Kumar and A. Srivastava, ``Enhanced performance of FSO link using OFDM and comparison with traditional TDM-FSO link," in \emph{Proc. IEEE Int. Broadband Photonics Conf.}, 2015, pp. 65-70.

\bibitem{FSO 2}
M. M. Tawfik, M. F. A. Sree, M. Abaza, and H. H. M. Ghouz, ``Inter-satellite optical wireless communication (IsOWC) system analysis for optimizing performance between GEO and LEO satellites," in \emph{Proc. Int. Telecommun. Conf.}, 2021, pp. 1-4.

\bibitem{computing energy consumption}
J. Wang, D. Feng, S. Zhang, A. Liu, and X. -G. Xia, ``Joint computation offloading and resource allocation for MEC-enabled IoT systems with imperfect CSI," \emph{IEEE Internet Things J.}, vol. 8, no. 5, pp. 3462-3475, Mar. 2021.

\bibitem{energy coefficient}
Y. Wang, M. Sheng, X. Wang, L. Wang, and J. Li, ``Mobile-edge computing: Partial computation offloading using dynamic voltage scaling," \emph{IEEE Trans. Commun.}, vol. 64, no. 10, pp. 4268-4282, Oct. 2016.

\bibitem{satellite network}
S. Zhang, G. Cui, Y. Long, and W. Wang, ``Joint computing and communication resource allocation for satellite communication networks with edge computing," \emph{China Commun.}, vol. 18, no. 7, pp. 236-252, Jul. 2021.

\bibitem{nphard}
J. Zhang, W. Xia, F. Yan, and L. Shen, ``Joint computation offloading and resource allocation optimization in heterogeneous networks with mobile edge computing," \emph{IEEE Access}, vol. 6, pp. 19324-19337, Apr. 2018.

\bibitem{AO}
J. C. Bezdek and R. J. Hathaway, ``Convergence of alternating optimization," \emph{Neural, Parallel \& Scientific Computations},
vol. 11, no. 4, pp. 351-368, Dec. 2003.

\bibitem{BnB}
J. Clausen, ``Branch and bound algorithms-principles and examples," \emph{Department of Computer Science, University of Copenhagen}, pp. 1-30, 1999.

\bibitem{CVX}
M. Grant and S. Boyd, \emph{CVX: Matlab Software for Disciplined Convex Programming}, [Online]: http://cvxr.com/cvx, Sep. 2013.

\bibitem{Convergence analyse}
V. A. Zorich and O. Paniagua, \emph{Mathematical analysis II}. Berlin: Springer, 2016.

\bibitem{IPM}
A. Ben-Tal and A. Nemirovski, \emph{Lectures on Modern Convex Optimization: Analysis, Algorithms, and Engineering Applications} (MPS-SIAM Series on Optimization). Philadelphia, PA, USA: SIAM, 2001.

\bibitem{complexity analysis}
K. Wang, A. M. So, T. Chang, W. Ma, and C. Chi, ``Outage constrained robust transmit optimization for multiuser MISO downlinks: Tractable approximations by conic optimization," \emph{IEEE Trans. Signal Process.}, vol. 62, no. 21, pp. 5690-5705, Nov. 2014.

\bibitem{Walker Delta1}
A. Al-Hourani, ``Session duration between handovers in dense LEO satellite networks," \emph{IEEE Wireless Commun. Lett.}, vol. 10, no. 12, pp. 2810-2814, Dec. 2021.

\bibitem{Walker Delta2}
C. Turner and R. T. Rajan, ``performance bounds for cooperative localisation in the starlink network," [Online]:https://arxiv.org/abs/2207.04691, 2022.

\bibitem{tau}
T. X. Tran and D. Pompili, ``Joint task offloading and resource allocation for multi-server mobile-edge computing networks," \emph{IEEE Trans. Veh. Technol.}, vol. 68, no. 1, pp. 856-868, Jan. 2019.

\bibitem{ZFBF}
M. Lin, Z. Lin, W. -P. Zhu, and J. -B. Wang, ``Joint beamforming for secure communication in cognitive satellite terrestrial networks," \emph{IEEE J. Sel. Areas Commun.}, vol. 36, no. 5, pp. 1017-1029, May 2018.

\bibitem{HCO}
J. Bi, H. Yuan, S. Duanmu, M. Zhou, and A. Abusorrah, ``Energy-optimized partial computation offloading in mobile-edge computing with genetic simulated-annealing-based particle swarm optimization," \emph{IEEE Internet Things J.}, vol. 8, no. 5, pp. 3774-3785, Mar. 2021.

\end{thebibliography}
\end{document}